\documentclass[reprint,amsmath,amssymb,aps,prl,showpacs,floatfix,superscriptaddress]{revtex4-1}

\usepackage{graphicx}
\usepackage{bm}
\usepackage{multirow}
\usepackage{verbatim}
\usepackage{longtable}
\usepackage{color}
\usepackage{nccmath}
\usepackage{siunitx} 
\usepackage{hyperref}

\renewcommand{\i}{i}

\newcommand\abs[1]{\lvert#1\rvert}

\usepackage[usenames,dvipsnames]{xcolor}

\begin{document}

\title{Observation of the exciton Mott transition in the photoluminescence\\
of coupled quantum wells}

\author{Gabija Kir{\v{s}}ansk{\.{e}}}
\affiliation{Niels Bohr Institute, University of Copenhagen, Blegdamsvej 17, DK-2100 Copenhagen, Denmark}
\author{Petru Tighineanu}
\affiliation{Niels Bohr Institute, University of Copenhagen, Blegdamsvej 17, DK-2100 Copenhagen, Denmark}
\author{Rapha\"{e}l S.~Daveau}
\affiliation{Niels Bohr Institute, University of Copenhagen, Blegdamsvej 17, DK-2100 Copenhagen, Denmark}
\author{Javier Miguel-S{\'{a}}nchez}
\affiliation{Institute of Quantum Electronics, ETH Z\"{u}rich, CH-8093, Z\"{u}rich, Switzerland}
\author{Peter Lodahl}
\affiliation{Niels Bohr Institute, University of Copenhagen, Blegdamsvej 17, DK-2100 Copenhagen, Denmark}
\author{S{\o}ren Stobbe}
\affiliation{Niels Bohr Institute, University of Copenhagen, Blegdamsvej 17, DK-2100 Copenhagen, Denmark}

\begin{abstract}
Indirect excitons in coupled quantum wells have long radiative lifetimes and form a cold quasi-two-dimensional population suitable for studying collective quantum effects. Here we report the observation of the exciton Mott transition from an insulating (excitons) to a conducting (ionized electron-hole pairs) phase, which occurs gradually as a function of carrier density and temperature. The transition is inferred from spectral and time-resolved photoluminescence measurements around a carrier density of $2\times10^{10} \mathrm{cm}^{-2}$ and temperatures of 12--\SI{16}{\kelvin}. An externally applied electric field is employed to tune the dynamics of the transition via the quantum-confined Stark effect. Our results provide evidence of a gradual nature of the exciton Mott transition.
\end{abstract}

\pacs{73.21.Fg, 78.67.D, 71.35.Lk, 71.10.Hf} 
\maketitle 



\begin{figure*}
\begin{center}
\includegraphics[width=1.0\textwidth]{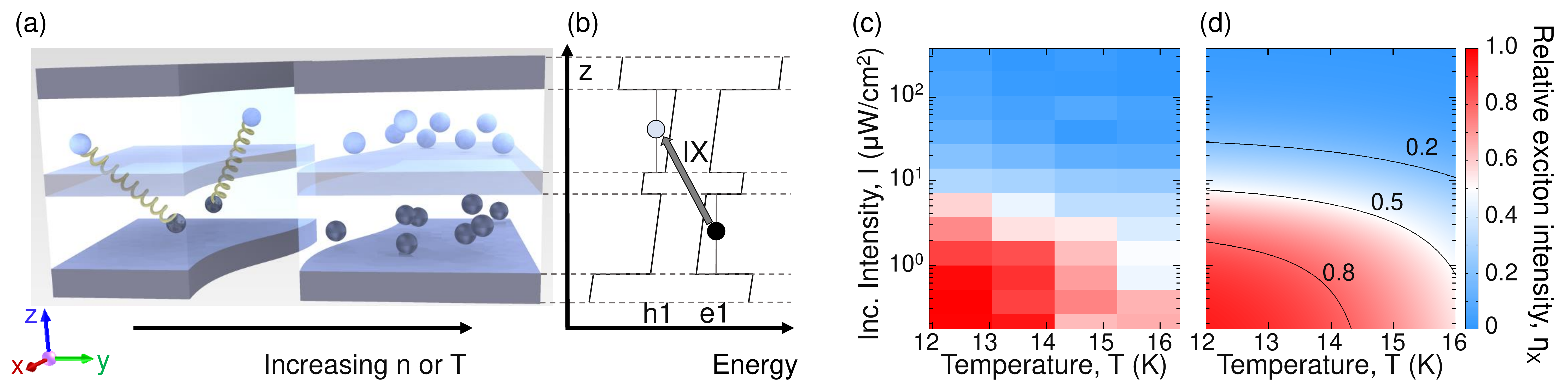}
\caption{The Mott transition of indirect excitons. (a) An exciton gas (left) is ionized into an EH(electron-hole)-plasma (right) as the exciton density or temperature is increased as sketched. The holes (electrons) are illustrated as light blue (black) spheres. (b) Band diagram and indirect transition of the CQWs. (c) The experimental phase diagram: the exciton intensity relative to the total intensity as a function of the incident (inc.) pumping intensity and the temperature. (d) The data in (c) are modelled by a statistical distribution function, see the main text for details. Contours of the relative exciton intensity of 0.2, 0.5, and 0.8 are indicated by the black lines.}
\label{fig1}
\end{center}
\end{figure*}

The Mott transition is a phase transition from an electrically insulating to a conducting state of matter predicted to occur in a system of correlated electrons~\cite{Mott1949}.  Over the past decades, it has been studied in different physical platforms such as semimetals, transition-metal compounds, doped semiconductors, organic salts, cold atomic gases, and superconductors~\cite{Neuenschwander1990, Zylbersztejn1975, Chernikov2015, Oliver1970, Sasaki1976, Limelette2003, Greiner2002, Poccia2015}. The transition may also happen in a population of interacting excitons, in which the increase of exciton density or temperature beyond a critical value induce a dissociation of bound excitons into a population of free electrons and holes~\cite{Mott1961}, cf.~Fig.~\ref{fig1}(a). The exciton Mott transition in two-dimensional systems has attracted particular interest and it was found to occur at electron-hole pair densities on the order of $10^{10}$--$10^{11} \mathrm{cm}^{-2}$ and temperatures around \SI{10}{\kelvin}~\cite{Finkelstein1995,Kaindl2003, Huber2005, Kappei2005, Stern2008, Rossbach2014}. So far, no consensus regarding the dynamics of the transition has been reached, in particular regarding the question of whether the ionization of excitons occurs abruptly~\cite{Ben-Taboude-Leon2003, Nikolaev2008} or gradually~\cite{Lozovik1996a,Koch2003,Manzke2012} as a function of the governing parameters, i.e., carrier density and temperature.

Indirect excitons (IXs) in coupled quantum wells (CQWs) have proven to be an attractive platform for uncovering fundamental quantum effects. The IX is composed of an electron and a hole residing in different quantum wells separated by a potential barrier, cf.~Figs.~\ref{fig1}(a,b). The small spatial overlap between the quasi-particles leads to a long radiative lifetime of the IX compared to a spatially direct exciton~\cite{Chen1987, Fisher1988,Hagn1995, Kim2005, Remeika2012}. Long-lived IXs can reach a thermodynamic equilibrium with the cold lattice allowing the study of coherent many-particle effects. Due to their rich and highly tunable properties, IXs have become an important platform for studying excitonic transport~\cite{Winbow2011, Butov1998}, cold gas condensation~\cite{Butov1994, Butov1998, Larionov2001, Butov2002a}, low-threshold lasing~\cite{Christmann2011}, terahertz generation~\cite{Luo1994}, and optical refrigeration~\cite{Daveaud2015}. The exciton Mott transition with IXs has so far been observed in photoluminescence (PL) spectroscopy from the spectral shape and energy shift of the excitonic peak~\cite{Burau2010, Stern2008}. The complexity of this multi-body phenomenon has, however, limited its understanding to mostly a qualitative level. Among many open questions, the dynamics of the transition is particularly unclear. In Refs.~\onlinecite{Kaindl2003, Huber2005, Kappei2005} the transition was found to occur gradually by THz or time-resolved photoluminescence spectroscopy, whereas Stern \emph{et al.}~\cite{Stern2008} reported an abrupt phase transition. Theoretical predictions have reached no consensus either~\cite{Ben-Taboude-Leon2003, Nikolaev2008,Lozovik1996a,Koch2003,Manzke2012}.

In the present paper we study the exciton Mott transition in InGaAs/GaAs/InGaAs 9/5/\SI{9}{\nano\meter} CQWs, cf.~Appendix A for further information about the sample. We find that the insulating and conducting populations emit at different energies and are easily resolved spectrally. We observe a gradual ionization of the excitonic peak with increasing excitation intensity and temperature as shown in Fig.~\ref{fig1}(c), which plots the ratio between the PL stemming from bound excitons and the total integrated PL.  The data presented in Fig.~\ref{fig1}(c) are modelled remarkably well by a two-dimensional distribution as shown in Fig.~\ref{fig1}(d). The model demonstrates that the transition observed in the experimental data behaves very close to the ideal scenario forecasted by Mott~\cite{Mott1949}, in which pumping intensity and temperature play a paramount yet independent role in the dynamics of the transition. Four characteristic parameters describing the transition are extracted. In the limit of low exciton density, the transition occurs at a temperature of $T_\textrm{M} = \SI{16.1}{\kelvin}$ and a rate of $\Gamma_T = \SI{1.4}{\kelvin}^{-1}$. Conversely, in the limit of low temperatures, the transition is observed at an excitation intensity of $I_\textrm{M} = \SI{7}{\watt/\centi\meter^2}$ and a rate of $\Gamma_I = \SI{2.7}{\centi\meter^{2}\per\watt}$. While previous studies of the exciton Mott transition were rather experimentally involved, requiring pump-probe techniques~\cite{Kaindl2003}, magnetic fields and special quasi-resonant pumping conditions~\cite{Stern2008}, our work provides a simple and direct demonstration of this fundamental many-body effect.

\begin{figure}
\begin{center}
\includegraphics[width=1\columnwidth]{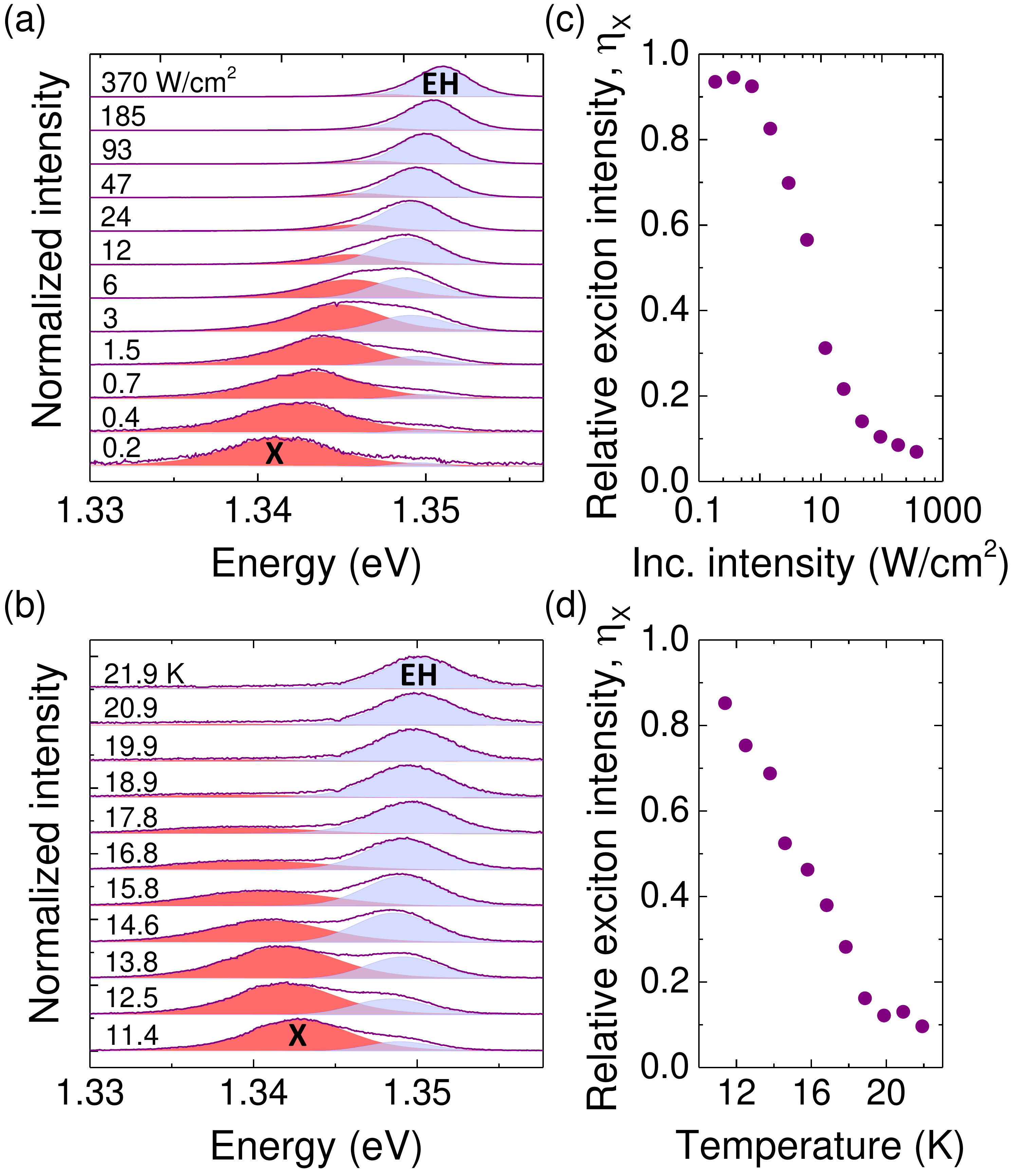}
\caption{Investigation of the exciton Mott transition as a function of excitation intensity and temperature. (a) Spectra of the IX taken at different incident intensity for a constant temperature of \SI{12.5}{\kelvin}. At low powers the PL stems mainly from excitons, which is gradually overtaken by the EH-plasma at larger carrier densities. (b) PL of the IX as a function of temperature for a constant excitation intensity of $\SI{1.5}{\watt/\centi\meter^2}$. As the temperature is increased, the excitons are ionized into free carriers. (c)--(d) Ratio between the PL stemming from the exciton resonance and the total PL extracted from (a) and (b).}
\label{fig2}
\end{center}
\end{figure}

Two key quantities govern the Mott transition --- excitation intensity and temperature. At low temperature and pumping power, the photo-generated carriers form hydrogen-like bonds due to their electrostatic attraction, i.e., excitons, which is reproduced in the data of Fig.~\ref{fig2}(a). Increasing the pumping power at a fixed temperature results in more excitons being ionized, which screen the electrostatic interaction between the remaining excitons, thus facilitating the ionization of more excitons~\cite{Rice1978}. This is described as a gradual enhancement of the plasma contribution, cf.~Fig.~\ref{fig2}(c). A continuous blueshift of the emission is observed with increasing pumping power because the electric field in the CQWs is screened by the indirect charge carriers.

Exciton ionization is also expected to occur with increasing temperature at a fixed pumping power. At low temperatures, the excitonic gas has a lower thermal energy than the typical binding energy of the IXs~\cite{Szymanska2003, Sivalertporn2012}. With increasing temperature, free electrons and holes facilitate the ionization of the excitons until the Coulomb bonds are broken. This can be seen in our data as a continuous decrease of the excitonic signal with temperature, cf.~Figs.~\ref{fig2}(b) and (d). The exciton PL redshifts with increasing temperature in accordance to Varshni's law, whereas the behavior of the plasma peak is somewhat peculiar because its spectral position is almost independent of temperature. The latter may be a consequence of a subtle interplay between the blueshift induced by screening effects and the redshift of Varshni's law. We model the spectrum with a sum of two Voigt functions, which take into account homogeneous and inhomogeneous broadening mechanisms~\cite{Humlicek1993}. The resulting distributions are separated by 3.5--\SI{8}{\milli\electronvolt}, which is comparable to the IX binding energy $E_b = \SI{3.8}{\milli\electronvolt}$ calculated using the numerical method presented in Ref.\ \onlinecite{Sivalertporn2012}. Other mechanisms, such as an interplay of bound and free excitons, are ruled out, see Appendix C.

We have recorded the dynamics of the Mott transition for several excitation intensities $I$ and temperatures $T$. For every phase-space configuration, the relative exciton intensity $\eta_\textrm{X}(I,T)$ is computed as the ratio between the integrated exciton PL $I_\textrm{X}(I,T)$ and the total integrated PL $I=I_\textrm{X}+I_\textrm{EH}$. The resulting phase diagram is plotted in Fig.~\ref{fig1}(c), which shows a gradual transition across the entire phase space. To acquire a quantitative understanding about the transition, we model the data in Fig.~\ref{fig1}(c) with a phenomenological two-dimensional logistic distribution. The cumulative function $\eta_\textrm{X}$ takes the form
\begin{equation}
\eta_\textrm{X} = \left[ 1 + e^{\Gamma_I\left(I-I_\textrm{M}\right)} \right]^{-1} \left[ 1 + e^{\Gamma_T\left(T-T_\textrm{M}\right)} \right]^{-1},
\label{eq:logisticDistribution}
\end{equation}
where $I_\textrm{M}$, $\Gamma_I$, $T_\textrm{M}$, $\Gamma_T$ are four numerically adjusted parameters whose values have been reported above. The boundary of the transition corresponds to $\eta_\textrm{X}=0.5$ and is visualized in Fig.~\ref{fig1}(d). A good reduced $\chi^2$ of the fit of 1.24 is obtained. Based on the extracted parameters, \SI{80}{\percent} of the transition occurs over a temperature interval $\Delta T = \SI{3.1}{\kelvin}$, and an intensity interval $\Delta I = \SI{1.6}{\watt\centi\meter^{-2}}$. A fundamental property of the Mott transition is that the two governing parameters, $I$ and $T$, play an independent role in the dynamics of the transition. To check whether this property is reflected in the data, we generalize Eq.~(\ref{eq:logisticDistribution}) to include correlations by performing the substitution $\Gamma_I I\rightarrow (1-\xi)^{-1/2}(\Gamma_I I + \xi\Gamma_T T)$ and analogously for $\Gamma_T T$ in the probability distribution $\rho_\textrm{X}=\partial^2 \eta_\textrm{X}/\partial T\partial I$, where $\abs{\xi}<1$ is the correlation factor. The resulting fit reveals a negligible correlation $\xi<0.05$, which implies that $I$ and $T$ drive the transition independently and is in agreement with Mott's prediction.

The characteristic parameter of the Mott transition is the critical exciton density $n_c$ at which the phase transition occurs. We define $n_c$ as the total carrier density at which the amount of PL stemming from the exciton and EH-plasma peaks is the same. An equal intensity results in a roughly equal density of the two species, i.e., $\eta_\textrm{X} = 0.5$. The critical density $n_c$ is estimated using a plate-capacitor model~\cite{Schindler2008}, which relates the exciton energy shift $\Delta E$, the distance between the opposite charges $d$, and the exciton density as $n_c = \Delta E \epsilon_r\epsilon_0/(e^2d)$, where $e$ is the elementary charge, $\epsilon_r$ is the relative dielectric constant and $\epsilon_0$ is the vacuum permittivity. In the simplest approximation, $d$ can be taken as the distance between the centers of the two quantum wells $d\sim \SI{14}{\nano\meter}$. From the measured IX energy shift we obtain a critical exciton density of $n_c = 2.3\times 10^{10} \ \mathrm{cm}^{-2}$ at $T=\SI{12.5}{\kelvin}$. We note that, as discussed by Ben-Tabou de-Leon \emph{et al.},\cite{Ben-Taboude-Leon2001} the capacitor formula underestimates the exciton density by around $50\%$ at densities of the order of $10^{10} \ \mathrm{cm}^{-2}$. An additional procedure is therefore used to estimate $n_c$ based on a steady-state optical absorption of the CQWs~\cite{Cerne1996}. We employ the model to calculate the CQWs absorption coefficient $\alpha = 1.2\times 10^{4} \ \mathrm{cm}^{-1}$, which allows estimating the number of absorbed photons per time $N_\mathrm{abs}$ in the CQWs of thickness $L$ yielding $N_\mathrm{abs}=\frac{P_i}{\hbar\omega}(1-e^{-\alpha L})$, where $P_i$ is the pumping power and $\omega$ the corresponding frequency. The radiative lifetime of the IX $\tau_\textrm{IX}$ and the measured emission area $A_\textrm{eff}$ are used to obtain $n_c = \frac{P_i}{\hbar\omega}\frac{1}{A_\textrm{eff}}(1-e^{-\alpha L})\tau_\textrm{IX}$, which results in $n_c$ $\sim$ 1.2$\times 10^{10} \ \mathrm{cm}^{-2}$ at \SI{12.5}{\kelvin}. The two independent estimates of $n_c$ differ by a factor of about two, which is likely caused by the fact that excitons with high in-plane momentum are optically dark and are not captured by the linear absorption method. We therefore employ the value of $n_c$ calculated by the plate-capacitor model, which agrees well with estimates for GaAs/AlGaAs and InGaAs/GaAs QWs reported in literature~\cite{Kaindl2003, Huber2005, Kappei2005, Stern2008}.
\begin{figure}
\centering
    \includegraphics[width=1.0\columnwidth]{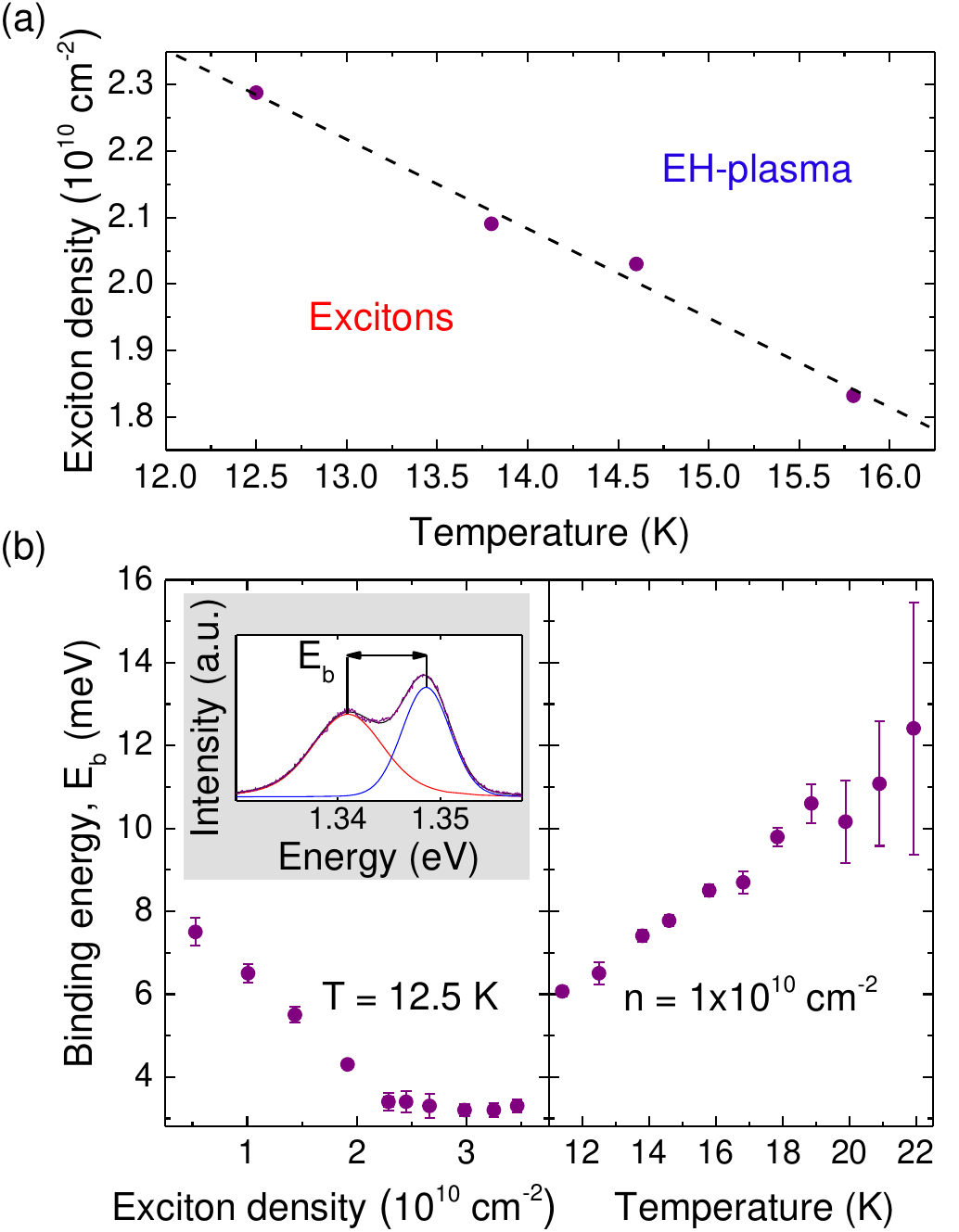}
\caption{(a) Density-temperature phase diagram of the exciton Mott transition. The critical exciton density $n_c$ is estimated for each temperature (data points). Two regions are distinguished where either the insulating excitons or the metallic plasma dominates. The black dashed line is a linear fit, $n_c = \beta T + \gamma$. (b) Binding energy of the IX across the Mott transition extracted from the data shown in Figs.~\ref{fig2}(a) and (b). As the carrier density $n$ is increased, the binding energy of the IX is reduced due to screening of the Coulomb interaction (left panel). At a carrier density of $1\times 10^{10}\si{\centi\meter^{-2}}$ and at \SI{12.5}{\kelvin}, the binding energy $E_b$ is \SI{6.5}{\milli\electronvolt} and becomes larger with increasing temperature (right panel). The error bars are calculated from the fitting errors of the centers of the exciton and EH-plasma distributions. Inset: PL spectrum fitted with a sum of two Voigt functions, where the distance between the peaks yields $E_b$.}
\label{fig3}
\end{figure}

To complete the picture of the exciton Mott transition, we provide a density-temperature phase diagram of the transition in Fig.~\ref{fig3}(a). The phase boundary is defined by the critical density $n_c$, which divides the diagram into two regions with exciton- or EH-plasma-dominant populations. The boundary can be well approximated by a linear dependence $n_c = \beta T + \gamma$ with $\beta = -\SI{0.137}{\centi\meter^{-2}\kelvin^{-1}}$ and $\gamma = 4\times 10^{10} \ \mathrm{cm}^{-2}$. The diagram partially agrees with the theoretical prediction from Refs.~\onlinecite{Nikolaev2008} and \onlinecite{Snoke2008}. In Ref.~\onlinecite{Snoke2008}, the relative fraction of excitons in a GaAs quantum well is calculated using the mass-action law in equilibrium and the static screening approximation, whereas Ref.~\onlinecite{Nikolaev2008} uses a Green-function formalism to calculate the phase diagram for CQWs. The picture is qualitatively similar in both works with an insulating state of excitons in the regime of low temperature and low density and a transition into the electron-hole plasma with an increase in one of the key parameters. Our results map out a rather restricted part of the suggested phase diagrams where the phase boundary is approximately linear. The limits of the phase boundary curve can be compared to theory using the model presented in Ref.~\onlinecite{Nikolaev2008}. For our CQWs, with a binding energy of the IX of $\SI{3.8}{\milli\electronvolt}$ and an exciton Bohr radius of $\SI{14}{\nano\meter}$, the exciton Mott transition is predicted to occur between $8.8$--$\SI{15.5}{\kelvin}$ and $1.5$--$5\times 10^{10}\si{\centi\meter^{-2}}$, which is in good agreement with our experimental phase diagram. Additionally, we compare the phase-boundary curve with the theoretical boundary presented in Ref.~\onlinecite{Snoke2008}, and find that at the density of $2 \times 10^{10}\si{\centi\meter^{-2}}$ the critical temperature for the Mott transition is 6--\SI{7}{\kelvin} lower than the values we measure~\cite{[{We compare the data to the linear part of the phase boundary curve for GaAs quantum wells presented by the dashed line in Fig.~6 in Ref~\cite{Snoke2008}}] empty}. Overall, the data are in good quantitative agreement with the theoretically predicted phase boundaries, but we do not observe a sharp boundary between the insulating and the metallic states. Instead, a gradual ionization of excitons occurs, which is consistent with the prediction of a continuous reduction of the exciton binding energy associated with the Mott transition~\cite{Haug1984,Zimmermann1988,Koch2003}.

The exciton binding energy can be extracted from the PL measurements. We extract the energy difference between the exciton and EH-plasma distributions from Figs.~\ref{fig2}(a) and (b), and denote it as the IX binding energy $E_b$, cf.~the inset in Fig.~\ref{fig3}(b). We find that $E_b$ is reduced as the carrier density increases, see left panel in Fig.~\ref{fig3}(b), as expected due to a carrier-induced screening of the Coulomb attraction between electrons and holes. We compare our data to the theoretical prediction of Ref.~\cite{Ben-Taboude-Leon2003}, where $E_b$ is found to decrease from $\SI{8}{\milli\electronvolt}$ to $2$--$\SI{4}{\milli\electronvolt}$ as the carrier density is increased from $0.5\times 10^{10}\si{\centi\meter^{-2}}$ to $3\times 10^{10}\si{\centi\meter^{-2}}$. This is in very good agreement with our results but we note that the temperature used in the calculation in Ref.~\cite{Ben-Taboude-Leon2003} is higher than in our experiment. A surprising feature is observed in the temperature dependence, cf.~right panel in Fig.~\ref{fig3}(b), where the binding energy is found to increase significantly. This may be caused by a reduced amount of screening due to the enhanced thermal energy of charge carriers as predicted by the theory in Ref.~\cite{Ben-Taboude-Leon2003}.

In summary, we have observed the excitonic Mott transition in CQWs and found a gradual ionization of the excitons with carrier density and temperature. We mapped the exciton-density-temperature phase diagram, which exhibits signatures of a second-order phase transition resulting in a phase coexistence~\cite{Lozovik1996a,Koch2003,Manzke2012}. Our analysis of the phase diagram led to the conclusion that temperature and carrier density trigger the Mott transition as independent parameters. We suggest that our observation of the Mott transition in PL experiments without magnetic fields or sophisticated techniques is made possible by the long exciton lifetime in the CQWs, which allows an efficient thermalization and the creation of a cold and coherent exciton population.

\begin{acknowledgments}

We gratefully acknowledge Ata{\c{c}} {\.{I}}mamo{\u{g}}lu for fruitful discussions. We thank the Lundbeck foundation, the Danish council for independent research (Natural Sciences and Technology and Production Sciences), and the European Research Council (ERC Consolidator Grant ALLQUANTUM) for the financial support.
\end{acknowledgments}

\section{Appendix A: Sample structure and experimental details}

\begin{figure*}[ht]
\begin{center}
\includegraphics[width=0.38\textwidth]{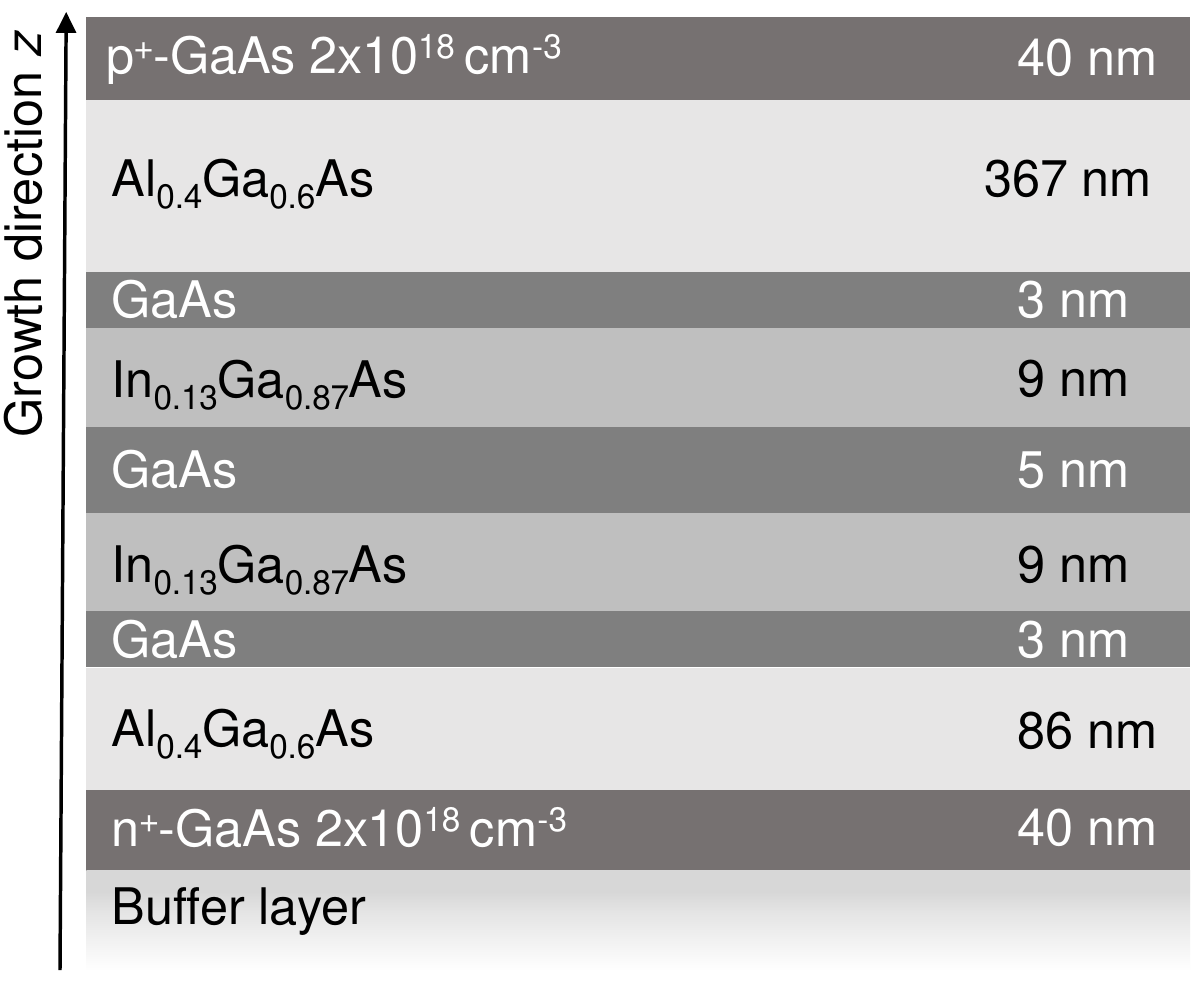}
\caption{Cross section of the wafer used in this work.}
\label{figSA1}
\end{center}
\end{figure*}

We study the exciton Mott transition in In$_{0.13}$Ga$_{0.87}$As/GaAs/In$_{0.13}$Ga$_{0.87}$As 9/5/\SI{9}{\nano\meter} CQWs grown by molecular-beam epitaxy and embedded in the intrinsic region of a p-i-n diode. The sample structure is shown in Fig.~\ref{figSA1}. Ohmic contacts are deposited on the doped GaAs layers allowing for electric-field application. For optical measurements, the sample is cooled to cryogenic temperatures. The temperature is measured by a calibrated Cernox sensor with an uncertainty of \SI{4}{\milli\kelvin} mounted next to the sample on the base plate and thermally anchored to avoid temperature offset due to heat load. Generally, the Cernox sensor provides a more accurate measurement of the sample temperature than measurements with the built-in sensor, which is usually mounted close to the cold finger, but does not capture additional deviations of the actual sample temperature due to laser heating and different thermal contact to the base plate.

For the optical measurements, a continuous-wave Ti:sapphire laser beam tuned to a wavelength of \SI{850}{\nano\meter} is focused to a spot of $\SI{13.5}{\micro\meter}^2$ via an objective of NA $= 0.25$, which is used to excite the sample from the top. This corresponds to excitation into a quasi-continuum of confined states of the CQWs. The PL from the CQWs is collected through the same objective, guided in free space to a spectrometer with a resolution of \SI{50}{\pico\meter} and detected by a cooled charge-coupled-device camera. The IX emission area is measured by exciting the sample at a constant position and collecting the PL from many different positions along a line, defined by a galvanometer mirror deflection. A picosecond pulsed diode laser with a repetition rate of \SI{2.5}{\mega\hertz} at \SI{785}{\nano\meter} is used for the time-resolved measurements. The signal is detected by an avalanche silicon photodiode.

\section{Appendix B: Direct and indirect transitions}
\label{sec:III}

\begin{figure*}[tp]
\begin{center}
\includegraphics[width=1.0\textwidth]{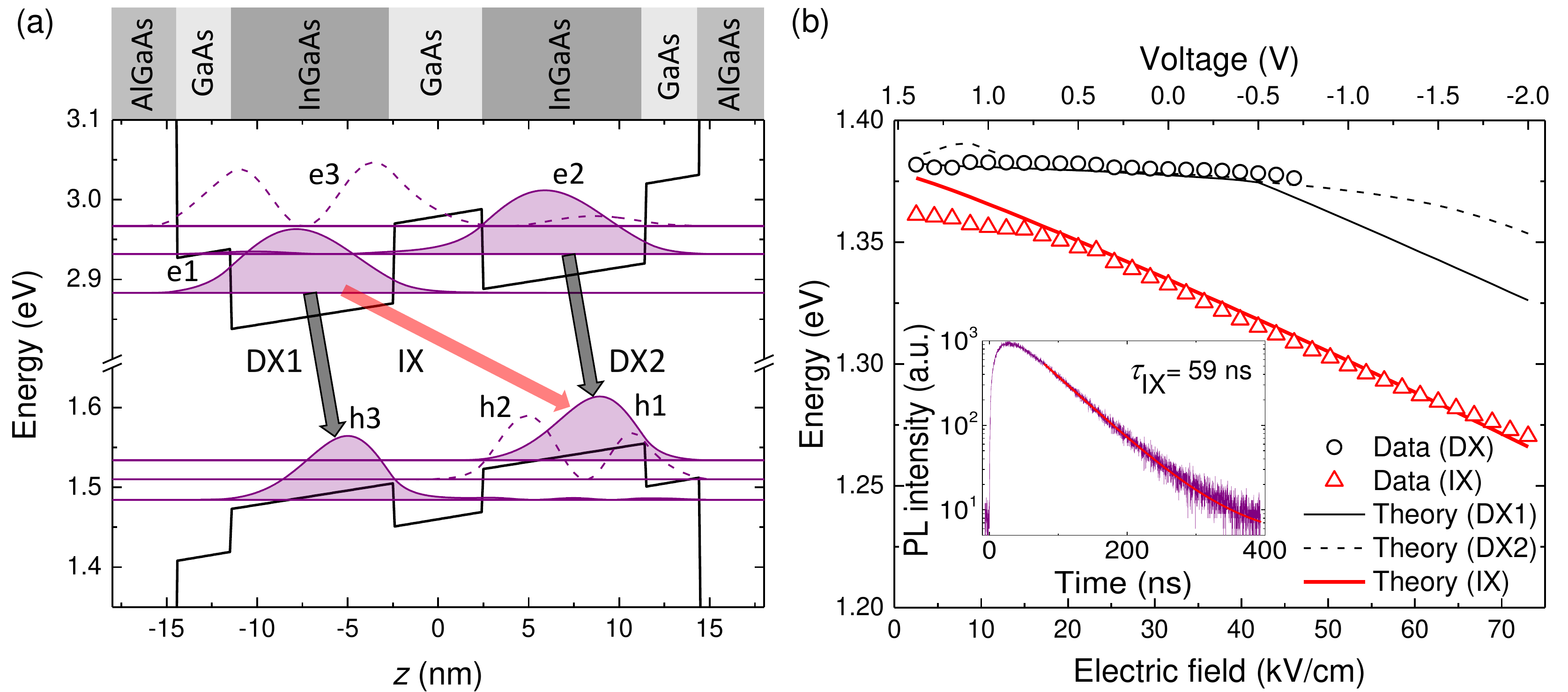}
\caption{Energy structure and the relevant optical transitions of the CQWs. (a) The band diagram at an external bias voltage of \SI{19}{\kilo\volt/\centi\meter} at which the Mott transition is observed. The first three electron (e1, e2, e3) and hole (h1, h2, h3) energy levels and squared wavefunctions are shown. The indirect-exciton PL originates from the e1 $\rightarrow$ h1 (red arrow) transition, while the direct excitons originate from e1 $\rightarrow$ h3 and e2 $\rightarrow$ h1 (black arrows). (b) Measured emission energy of DX2 (black circles) and the IX (red triangles) versus electric field at an excitation intensity of $\SI{545}{\watt/\centi\meter^2}$, which agrees very well with the theoretical predictions (solid lines). Inset: the IX decay dynamics recorded at a pumping intensity of $\SI{319}{\watt/\centi\meter^2}$ and an electric field of \SI{19}{\kilo\volt/\centi\meter}.}
\label{figSA2}
\end{center}
\end{figure*}

Here we present the basic electronic and optical properties of the CQWs. The band diagram is evaluated numerically using a routine that solves the single-particle effective-mass Schr\"{o}dinger equation with a tunneling resonance technique~\cite{Miller1985} for an arbitrary potential distribution (the band parameters are taken from Ref.~\onlinecite{Vurgaftman2001}). Radiative and non-radiative lifetimes as well as absorption coefficients are calculated using Fermi's golden rule. It is sufficient to account for the first three eigenstates of electrons (e1, e2, e3) and heavy holes (h1, h2, h3), cf.~Fig.~\ref{figSA2}(a). The electric field renders the ground-state wavefunctions of electrons, e1, and holes, h1, localized in opposite quantum wells, which results in the formation of a spatially indirect transition. Another indirect transition e1~$\rightarrow$~h2 is optically weak and is not discussed further. The relevant excited states of the CQWs are the spatially direct transitions e1-h3 and e2-h1.

To identify the direct and indirect transitions, we study the PL as a function of the applied bias voltage, cf.~Fig.~\ref{figSA2}(b). The spectral position of the direct transitions is nearly independent of the bias as expected for a DX with negligible static dipole moment~\cite{Sivalertporn2012}. Above \SI{40}{\kilo\volt/\centi\meter} the DX becomes indirect and responsive to the electric field. The IX has a large permanent dipole moment determined by the distance between the quantum wells of about \SI{14}{\nano\meter}. The recombination energy of the IX depends linearly on the applied field and can be tuned significantly~\cite{Chen1987, Andrews1988, Kim2005}. We compare the measured IX and DX peak shifts with our theoretical model and find good agreement, cf.~Fig.~\ref{figSA2}(b). The discrepancy at large forward bias might be caused by the resistance of the ohmic contacts, which results in a parasitic voltage drop in the contacts. Time-resolved measurements confirm that IXs are long lived with a lifetime $\tau_\textrm{IX} = \SI{59}{\nano\second}$, cf.~Fig.~\ref{figSA2}(b). Under these conditions, the radiative lifetime of the IX is much longer than the typical IX thermalization time of a few nanoseconds~\cite{Butov2001}. The IXs therefore establish a thermodynamic equilibrium with the cold crystal lattice, which allows the study of low-temperature collective effects such as the Mott transition.

\begin{figure}
\centering
    \includegraphics[width=0.45\textwidth]{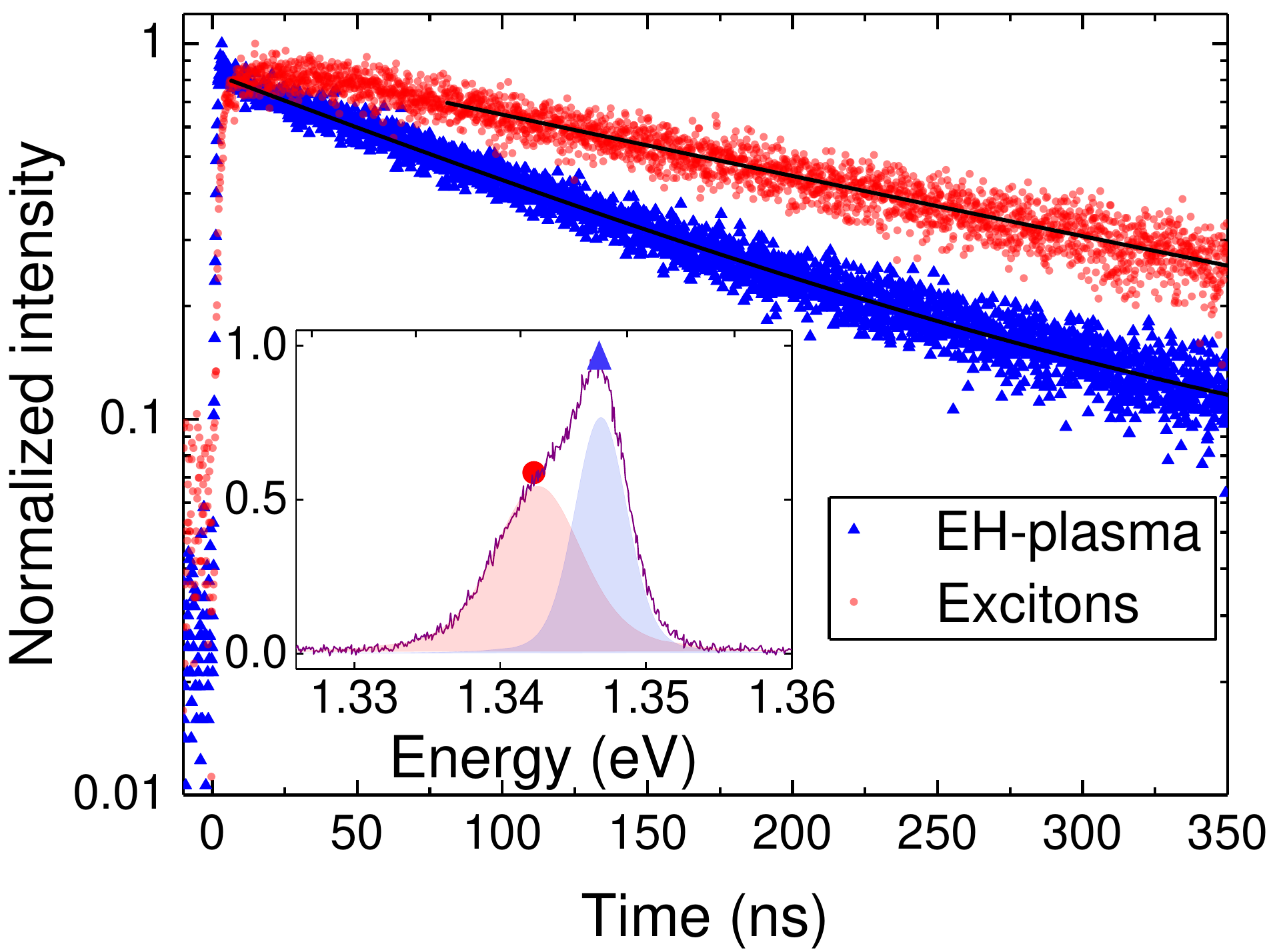}
\caption{Time-resolved decay dynamics of excitons (red dots) and EH-plasma (blue triangles) taken at a pumping intensity of $\SI{1}{\watt/\centi\meter}^2$ under \SI{19}{\kilo\volt/\centi\meter} applied bias. The curves are fitted well by a single (excitons) or double (EH-plasma) exponent (black solid lines). Inset: the corresponding spectrum of the indirect transition under pulsed excitation. The red dot and blue triangle denote the two spectral positions probed by the time-resolved measurement.}
\label{figSA3}
\end{figure}
We perform time-resolved measurements on the two species populating the indirect transition to study their decay dynamics. The measurements are recorded at two spectral positions corresponding to the central emission frequency of the two populations, cf.~the inset of Fig.~\ref{figSA3}. We observe a faster decay of electron-hole plasma at a rate of \SI{6.6}{\micro\second^{-1}}, whereas the excitons decay at a rate of \SI{4.6}{\micro\second^{-1}}. A similar behavior of the mobile excitons and the free carriers has been observed in previous work on GaAs and InGaAs quantum wells, but with the lifetime of both types of population on the order of a nanosecond~\cite{Amo2007, Kappei2005}.

\section{Appendix C: Ruling out other possible mechanisms}

\begin{figure*}[ht]
\begin{center}
\includegraphics[width=1.0\textwidth]{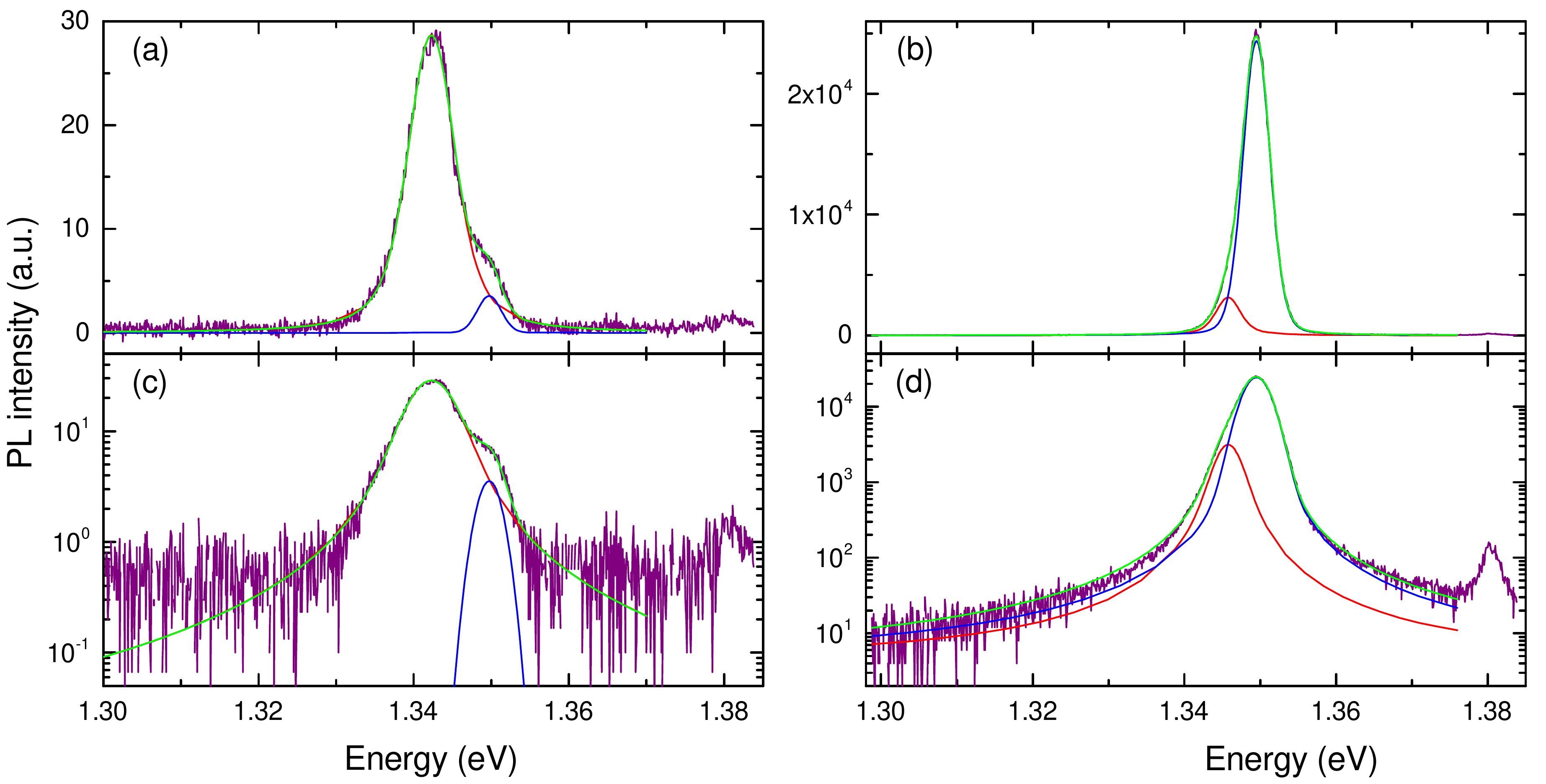}
\caption{PL spectra of CQWs (purple line) acquired at an electric field of \SI{19}{\kilo\volt/\centi\meter} shown on a linear and semilogarithmic scale for an excitation intensity of $\SI{0.7}{\watt/\centi\meter^2}$ (a), (c) and $\SI{47}{\watt/\centi\meter^2}$ (b), (d), respectively. The IX peak is fitted with a sum of two Voigt functions (green line), where the excitonic (EH-plasma) contribution is depicted in red (blue). The peak at high energies stems from the direct exciton.}
\label{figSA4}
\end{center}
\end{figure*}

Since the excitonic Mott transition occurs in a solid-state environment, which hosts a plethora of physical phenomena, it is essential to rule out alternative mechanisms that could have explained our data. A common practice in literature is to analyze PL spectra on a semilogarithmic scale~\cite{Rossbach2014,Shahmohammadi2014}. The signature of EH-plasma in direct quantum wells is considered an exponentially decaying high-energy tail of the PL peak. Our measurements show a much more pronounced plasma feature than in direct QWs, which, unlike in direct quantum wells, is probably caused by an efficient thermalization of charge carriers. We show power-dependent PL spectra of CQWs and the corresponding fits in Fig.~\ref{figSA4}. The EH-plasma contribution is significant at all excitation intensities, and its lineshape is well reproduced by a Voigt function, see Figs.~\ref{figSA4}(a) and (c). The plasma dominates the PL emission at higher intensities, c.f.~Figs.~\ref{figSA4}(b) and (d).

A mechanism resulting in a somewhat similar PL spectrum is the interplay of localized and free excitons~\cite{High2009, Sugakov2014}. The localized excitons are caused by doping impurities or fluctuations in the quantum-well thickness~\cite{Hoger1984, Leosson2000,High2009}, which form random localized traps for the excitons. In typical quality samples these traps are saturated with relative ease due to Pauli blocking~\cite{High2009, Sugakov2014} at carrier densities of the order of $10^{9}\si{\centi\meter^{-2}}$ below \SI{2}{\kelvin}~\cite{Larionov2002, Dremin2002}, and at even lower densities at higher temperatures~\cite{Cui1996}. In our measurements, however, the low-energy peak does not reach saturation even at carrier densities of the order of $10^{11}\si{\centi\meter^{-2}}$. Additionally, sharp features are expected in the PL spectrum due to the spectrally narrow density of states of the trapping sites~\cite{High2009, Leosson2000}. In our case, no such features or changes in the structure are present in the spectrum even at the smallest excitation intensities corresponding to the exciton density of $5.2 \times 10^{9}\si{\centi\meter^{-2}}$.

A possible reason for a doublet structure in the spectrum of the indirect transition could be related to quantum wells of different thickness. Due to a finite growth accuracy, the layer thickness might fluctuate by an atomic monolayer (ML). We calculate that a fluctuation of 1 ML results in an energy shift of around \SI{1.5}{\milli\electronvolt}, and 1 ML deviation in the barrier thickness $b$ induces a shift of less than \SI{1}{\milli\electronvolt}. Since the energy difference between the two resonances of the indirect transition varies between 6--\SI{10}{\milli\electronvolt}, we conclude that this effect cannot explain our data.

Other possible mechanisms like thermally activated high-energy states are excluded, since the energy difference between the ground state and the first excited exciton state is much larger than the spectral separation of the indirect doublet.


\begin{thebibliography}{59}%
\makeatletter
\providecommand \@ifxundefined [1]{%
 \@ifx{#1\undefined}
}%
\providecommand \@ifnum [1]{%
 \ifnum #1\expandafter \@firstoftwo
 \else \expandafter \@secondoftwo
 \fi
}%
\providecommand \@ifx [1]{%
 \ifx #1\expandafter \@firstoftwo
 \else \expandafter \@secondoftwo
 \fi
}%
\providecommand \natexlab [1]{#1}%
\providecommand \enquote  [1]{``#1''}%
\providecommand \bibnamefont  [1]{#1}%
\providecommand \bibfnamefont [1]{#1}%
\providecommand \citenamefont [1]{#1}%
\providecommand \href@noop [0]{\@secondoftwo}%
\providecommand \href [0]{\begingroup \@sanitize@url \@href}%
\providecommand \@href[1]{\@@startlink{#1}\@@href}%
\providecommand \@@href[1]{\endgroup#1\@@endlink}%
\providecommand \@sanitize@url [0]{\catcode `\\12\catcode `\$12\catcode
  `\&12\catcode `\#12\catcode `\^12\catcode `\_12\catcode `\%12\relax}%
\providecommand \@@startlink[1]{}%
\providecommand \@@endlink[0]{}%
\providecommand \url  [0]{\begingroup\@sanitize@url \@url }%
\providecommand \@url [1]{\endgroup\@href {#1}{\urlprefix }}%
\providecommand \urlprefix  [0]{URL }%
\providecommand \Eprint [0]{\href }%
\providecommand \doibase [0]{http://dx.doi.org/}%
\providecommand \selectlanguage [0]{\@gobble}%
\providecommand \bibinfo  [0]{\@secondoftwo}%
\providecommand \bibfield  [0]{\@secondoftwo}%
\providecommand \translation [1]{[#1]}%
\providecommand \BibitemOpen [0]{}%
\providecommand \bibitemStop [0]{}%
\providecommand \bibitemNoStop [0]{.\EOS\space}%
\providecommand \EOS [0]{\spacefactor3000\relax}%
\providecommand \BibitemShut  [1]{\csname bibitem#1\endcsname}%
\let\auto@bib@innerbib\@empty
\bibitem [{\citenamefont {Mott}(1949)}]{Mott1949}%
  \BibitemOpen
  \bibfield  {author} {\bibinfo {author} {\bibfnamefont {N.~F.}\ \bibnamefont
  {Mott}},\ }\href {\doibase 10.1088/0370-1298/62/7/303} {\bibfield  {journal}
  {\bibinfo  {journal} {Proc. Phys. Soc. Sect. A}\ }\textbf {\bibinfo {volume}
  {62}},\ \bibinfo {pages} {416} (\bibinfo {year} {1949})}\BibitemShut
  {NoStop}%
\bibitem [{\citenamefont {Neuenschwander}\ and\ \citenamefont
  {Wachter}(1990)}]{Neuenschwander1990}%
  \BibitemOpen
  \bibfield  {author} {\bibinfo {author} {\bibfnamefont {J.}~\bibnamefont
  {Neuenschwander}}\ and\ \bibinfo {author} {\bibfnamefont {P.}~\bibnamefont
  {Wachter}},\ }\href {\doibase 10.1103/PhysRevB.41.12693} {\bibfield
  {journal} {\bibinfo  {journal} {Phys. Rev. B}\ }\textbf {\bibinfo {volume}
  {41}},\ \bibinfo {pages} {12693} (\bibinfo {year} {1990})}\BibitemShut
  {NoStop}%
\bibitem [{\citenamefont {Zylbersztejn}\ and\ \citenamefont
  {Mott}(1975)}]{Zylbersztejn1975}%
  \BibitemOpen
  \bibfield  {author} {\bibinfo {author} {\bibfnamefont {A.}~\bibnamefont
  {Zylbersztejn}}\ and\ \bibinfo {author} {\bibfnamefont {N.~F.}\ \bibnamefont
  {Mott}},\ }\href {\doibase 10.1103/PhysRevB.11.4383} {\bibfield  {journal}
  {\bibinfo  {journal} {Phys. Rev. B}\ }\textbf {\bibinfo {volume} {11}},\
  \bibinfo {pages} {4383} (\bibinfo {year} {1975})}\BibitemShut {NoStop}%
\bibitem [{\citenamefont {Chernikov}\ \emph {et~al.}(2015)\citenamefont
  {Chernikov}, \citenamefont {Ruppert}, \citenamefont {Hill}, \citenamefont
  {Rigosi},\ and\ \citenamefont {Heinz}}]{Chernikov2015}%
  \BibitemOpen
  \bibfield  {author} {\bibinfo {author} {\bibfnamefont {A.}~\bibnamefont
  {Chernikov}}, \bibinfo {author} {\bibfnamefont {C.}~\bibnamefont {Ruppert}},
  \bibinfo {author} {\bibfnamefont {H.~M.}\ \bibnamefont {Hill}}, \bibinfo
  {author} {\bibfnamefont {A.~F.}\ \bibnamefont {Rigosi}}, \ and\ \bibinfo
  {author} {\bibfnamefont {T.~F.}\ \bibnamefont {Heinz}},\ }\href {\doibase
  10.1038/nphoton.2015.104} {\bibfield  {journal} {\bibinfo  {journal} {Nat.
  Photonics}\ }\textbf {\bibinfo {volume} {9}},\ \bibinfo {pages} {466}
  (\bibinfo {year} {2015})}\BibitemShut {NoStop}%
\bibitem [{\citenamefont {Oliver}\ \emph {et~al.}(1970)\citenamefont {Oliver},
  \citenamefont {Kafalas}, \citenamefont {Dimmock},\ and\ \citenamefont
  {Reed}}]{Oliver1970}%
  \BibitemOpen
  \bibfield  {author} {\bibinfo {author} {\bibfnamefont {M.~R.}\ \bibnamefont
  {Oliver}}, \bibinfo {author} {\bibfnamefont {J.~A.}\ \bibnamefont {Kafalas}},
  \bibinfo {author} {\bibfnamefont {J.~O.}\ \bibnamefont {Dimmock}}, \ and\
  \bibinfo {author} {\bibfnamefont {T.~B.}\ \bibnamefont {Reed}},\ }\href
  {\doibase http://dx.doi.org/10.1103/PhysRevLett.24.1064} {\bibfield
  {journal} {\bibinfo  {journal} {Phys. Rev. Lett.}\ }\textbf {\bibinfo
  {volume} {24}},\ \bibinfo {pages} {1064} (\bibinfo {year}
  {1970})}\BibitemShut {NoStop}%
\bibitem [{\citenamefont {Sasaki}(1976)}]{Sasaki1976}%
  \BibitemOpen
  \bibfield  {author} {\bibinfo {author} {\bibfnamefont {W.}~\bibnamefont
  {Sasaki}},\ }\href {\doibase 10.1051/jphyscol:1976454} {\bibfield  {journal}
  {\bibinfo  {journal} {Le J. Phys. Colloq.}\ }\textbf {\bibinfo {volume}
  {37}},\ \bibinfo {pages} {C4} (\bibinfo {year} {1976})}\BibitemShut {NoStop}%
\bibitem [{\citenamefont {Limelette}\ \emph {et~al.}(2003)\citenamefont
  {Limelette}, \citenamefont {Wzietek}, \citenamefont {Florens}, \citenamefont
  {Georges}, \citenamefont {Costi}, \citenamefont {Pasquier}, \citenamefont
  {J\'{e}rome}, \citenamefont {M\'{e}zi\`{e}re},\ and\ \citenamefont
  {Batail}}]{Limelette2003}%
  \BibitemOpen
  \bibfield  {author} {\bibinfo {author} {\bibfnamefont {P.}~\bibnamefont
  {Limelette}}, \bibinfo {author} {\bibfnamefont {P.}~\bibnamefont {Wzietek}},
  \bibinfo {author} {\bibfnamefont {S.}~\bibnamefont {Florens}}, \bibinfo
  {author} {\bibfnamefont {A.}~\bibnamefont {Georges}}, \bibinfo {author}
  {\bibfnamefont {T.~A.}\ \bibnamefont {Costi}}, \bibinfo {author}
  {\bibfnamefont {C.}~\bibnamefont {Pasquier}}, \bibinfo {author}
  {\bibfnamefont {D.}~\bibnamefont {J\'{e}rome}}, \bibinfo {author}
  {\bibfnamefont {C.}~\bibnamefont {M\'{e}zi\`{e}re}}, \ and\ \bibinfo {author}
  {\bibfnamefont {P.}~\bibnamefont {Batail}},\ }\href {\doibase
  10.1103/PhysRevLett.91.016401} {\bibfield  {journal} {\bibinfo  {journal}
  {Phys. Rev. Lett.}\ }\textbf {\bibinfo {volume} {91}},\ \bibinfo {pages}
  {016401} (\bibinfo {year} {2003})}\BibitemShut {NoStop}%
\bibitem [{\citenamefont {Greiner}\ \emph {et~al.}(2002)\citenamefont
  {Greiner}, \citenamefont {Mandel}, \citenamefont {Esslinger}, \citenamefont
  {H\"{a}nsch},\ and\ \citenamefont {Bloch}}]{Greiner2002}%
  \BibitemOpen
  \bibfield  {author} {\bibinfo {author} {\bibfnamefont {M.}~\bibnamefont
  {Greiner}}, \bibinfo {author} {\bibfnamefont {O.}~\bibnamefont {Mandel}},
  \bibinfo {author} {\bibfnamefont {T.}~\bibnamefont {Esslinger}}, \bibinfo
  {author} {\bibfnamefont {T.~W.}\ \bibnamefont {H\"{a}nsch}}, \ and\ \bibinfo
  {author} {\bibfnamefont {I.}~\bibnamefont {Bloch}},\ }\href {\doibase
  10.1038/415039a} {\bibfield  {journal} {\bibinfo  {journal} {Nature}\
  }\textbf {\bibinfo {volume} {415}},\ \bibinfo {pages} {39} (\bibinfo {year}
  {2002})}\BibitemShut {NoStop}%
\bibitem [{\citenamefont {Poccia}\ \emph {et~al.}(2015)\citenamefont {Poccia},
  \citenamefont {Baturina}, \citenamefont {Coneri}, \citenamefont {Molenaar},
  \citenamefont {Wang}, \citenamefont {Bianconi}, \citenamefont {Brinkman},
  \citenamefont {Hilgenkamp}, \citenamefont {Golubov},\ and\ \citenamefont
  {Vinokur}}]{Poccia2015}%
  \BibitemOpen
  \bibfield  {author} {\bibinfo {author} {\bibfnamefont {N.}~\bibnamefont
  {Poccia}}, \bibinfo {author} {\bibfnamefont {T.~I.}\ \bibnamefont
  {Baturina}}, \bibinfo {author} {\bibfnamefont {F.}~\bibnamefont {Coneri}},
  \bibinfo {author} {\bibfnamefont {C.~G.}\ \bibnamefont {Molenaar}}, \bibinfo
  {author} {\bibfnamefont {X.~R.}\ \bibnamefont {Wang}}, \bibinfo {author}
  {\bibfnamefont {G.}~\bibnamefont {Bianconi}}, \bibinfo {author}
  {\bibfnamefont {A.}~\bibnamefont {Brinkman}}, \bibinfo {author}
  {\bibfnamefont {H.}~\bibnamefont {Hilgenkamp}}, \bibinfo {author}
  {\bibfnamefont {A.~A.}\ \bibnamefont {Golubov}}, \ and\ \bibinfo {author}
  {\bibfnamefont {V.~M.}\ \bibnamefont {Vinokur}},\ }\href {\doibase
  10.1126/science.1260507} {\bibfield  {journal} {\bibinfo  {journal}
  {Science}\ }\textbf {\bibinfo {volume} {349}},\ \bibinfo {pages} {1202}
  (\bibinfo {year} {2015})}\BibitemShut {NoStop}%
\bibitem [{\citenamefont {Mott}(1961)}]{Mott1961}%
  \BibitemOpen
  \bibfield  {author} {\bibinfo {author} {\bibfnamefont {N.~F.}\ \bibnamefont
  {Mott}},\ }\href {\doibase 10.1080/14786436108243318} {\bibfield  {journal}
  {\bibinfo  {journal} {Philos. Mag.}\ }\textbf {\bibinfo {volume} {6}},\
  \bibinfo {pages} {287} (\bibinfo {year} {1961})}\BibitemShut {NoStop}%
\bibitem [{\citenamefont {Finkelstein}\ \emph {et~al.}(1995)\citenamefont
  {Finkelstein}, \citenamefont {Shtrikman},\ and\ \citenamefont
  {Bar-Joseph}}]{Finkelstein1995}%
  \BibitemOpen
  \bibfield  {author} {\bibinfo {author} {\bibfnamefont {G.}~\bibnamefont
  {Finkelstein}}, \bibinfo {author} {\bibfnamefont {H.}~\bibnamefont
  {Shtrikman}}, \ and\ \bibinfo {author} {\bibfnamefont {I.}~\bibnamefont
  {Bar-Joseph}},\ }\href {\doibase 10.1103/PhysRevLett.74.976} {\bibfield
  {journal} {\bibinfo  {journal} {Phys. Rev. Lett.}\ }\textbf {\bibinfo
  {volume} {74}},\ \bibinfo {pages} {976} (\bibinfo {year} {1995})}\BibitemShut
  {NoStop}%
\bibitem [{\citenamefont {Kaindl}\ \emph {et~al.}(2003)\citenamefont {Kaindl},
  \citenamefont {Carnahan}, \citenamefont {H\"{a}gele}, \citenamefont
  {L\"{o}venich},\ and\ \citenamefont {Chemla}}]{Kaindl2003}%
  \BibitemOpen
  \bibfield  {author} {\bibinfo {author} {\bibfnamefont {R.}~\bibnamefont
  {Kaindl}}, \bibinfo {author} {\bibfnamefont {M.~A.}\ \bibnamefont
  {Carnahan}}, \bibinfo {author} {\bibfnamefont {D.}~\bibnamefont
  {H\"{a}gele}}, \bibinfo {author} {\bibfnamefont {R.}~\bibnamefont
  {L\"{o}venich}}, \ and\ \bibinfo {author} {\bibfnamefont {D.~S.}\
  \bibnamefont {Chemla}},\ }\href {\doibase 10.1038/nature01714.1.} {\bibfield
  {journal} {\bibinfo  {journal} {Nature}\ }\textbf {\bibinfo {volume} {423}},\
  \bibinfo {pages} {734} (\bibinfo {year} {2003})}\BibitemShut {NoStop}%
\bibitem [{\citenamefont {Huber}\ \emph {et~al.}(2005)\citenamefont {Huber},
  \citenamefont {Kaindl}, \citenamefont {Schmid},\ and\ \citenamefont
  {Chemla}}]{Huber2005}%
  \BibitemOpen
  \bibfield  {author} {\bibinfo {author} {\bibfnamefont {R.}~\bibnamefont
  {Huber}}, \bibinfo {author} {\bibfnamefont {R.}~\bibnamefont {Kaindl}},
  \bibinfo {author} {\bibfnamefont {B.}~\bibnamefont {Schmid}}, \ and\ \bibinfo
  {author} {\bibfnamefont {D.}~\bibnamefont {Chemla}},\ }\href {\doibase
  10.1103/PhysRevB.72.161314} {\bibfield  {journal} {\bibinfo  {journal} {Phys.
  Rev. B}\ }\textbf {\bibinfo {volume} {72}},\ \bibinfo {pages} {161314}
  (\bibinfo {year} {2005})}\BibitemShut {NoStop}%
\bibitem [{\citenamefont {Kappei}\ \emph {et~al.}(2005)\citenamefont {Kappei},
  \citenamefont {Szczytko}, \citenamefont {Morier-Genoud},\ and\ \citenamefont
  {Deveaud}}]{Kappei2005}%
  \BibitemOpen
  \bibfield  {author} {\bibinfo {author} {\bibfnamefont {L.}~\bibnamefont
  {Kappei}}, \bibinfo {author} {\bibfnamefont {J.}~\bibnamefont {Szczytko}},
  \bibinfo {author} {\bibfnamefont {F.}~\bibnamefont {Morier-Genoud}}, \ and\
  \bibinfo {author} {\bibfnamefont {B.}~\bibnamefont {Deveaud}},\ }\href
  {\doibase 10.1103/PhysRevLett.94.147403} {\bibfield  {journal} {\bibinfo
  {journal} {Phys. Rev. Lett.}\ }\textbf {\bibinfo {volume} {94}},\ \bibinfo
  {pages} {147403} (\bibinfo {year} {2005})}\BibitemShut {NoStop}%
\bibitem [{\citenamefont {Stern}\ \emph {et~al.}(2008)\citenamefont {Stern},
  \citenamefont {Garmider}, \citenamefont {Umansky},\ and\ \citenamefont
  {Bar-Joseph}}]{Stern2008}%
  \BibitemOpen
  \bibfield  {author} {\bibinfo {author} {\bibfnamefont {M.}~\bibnamefont
  {Stern}}, \bibinfo {author} {\bibfnamefont {V.}~\bibnamefont {Garmider}},
  \bibinfo {author} {\bibfnamefont {V.}~\bibnamefont {Umansky}}, \ and\
  \bibinfo {author} {\bibfnamefont {I.}~\bibnamefont {Bar-Joseph}},\ }\href
  {\doibase 10.1103/PhysRevLett.100.256402} {\bibfield  {journal} {\bibinfo
  {journal} {Phys. Rev. Lett.}\ }\textbf {\bibinfo {volume} {100}},\ \bibinfo
  {pages} {256402} (\bibinfo {year} {2008})}\BibitemShut {NoStop}%
\bibitem [{\citenamefont {Rossbach}\ \emph {et~al.}(2014)\citenamefont
  {Rossbach}, \citenamefont {Levrat}, \citenamefont {Jacopin}, \citenamefont
  {Shahmohammadi}, \citenamefont {Carlin}, \citenamefont {Gani\`{e}re},
  \citenamefont {Butt\'{e}}, \citenamefont {Deveaud},\ and\ \citenamefont
  {Grandjean}}]{Rossbach2014}%
  \BibitemOpen
  \bibfield  {author} {\bibinfo {author} {\bibfnamefont {G.}~\bibnamefont
  {Rossbach}}, \bibinfo {author} {\bibfnamefont {J.}~\bibnamefont {Levrat}},
  \bibinfo {author} {\bibfnamefont {G.}~\bibnamefont {Jacopin}}, \bibinfo
  {author} {\bibfnamefont {M.}~\bibnamefont {Shahmohammadi}}, \bibinfo {author}
  {\bibfnamefont {J.-F.}\ \bibnamefont {Carlin}}, \bibinfo {author}
  {\bibfnamefont {J.-D.}\ \bibnamefont {Gani\`{e}re}}, \bibinfo {author}
  {\bibfnamefont {R.}~\bibnamefont {Butt\'{e}}}, \bibinfo {author}
  {\bibfnamefont {B.}~\bibnamefont {Deveaud}}, \ and\ \bibinfo {author}
  {\bibfnamefont {N.}~\bibnamefont {Grandjean}},\ }\href {\doibase
  10.1103/PhysRevB.90.201308} {\bibfield  {journal} {\bibinfo  {journal} {Phys.
  Rev. B}\ }\textbf {\bibinfo {volume} {90}},\ \bibinfo {pages} {201308}
  (\bibinfo {year} {2014})}\BibitemShut {NoStop}%
\bibitem [{\citenamefont {Ben-Tabou~de Leon}\ and\ \citenamefont
  {Laikhtman}(2003)}]{Ben-Taboude-Leon2003}%
  \BibitemOpen
  \bibfield  {author} {\bibinfo {author} {\bibfnamefont {S.}~\bibnamefont
  {Ben-Tabou~de Leon}}\ and\ \bibinfo {author} {\bibfnamefont {B.}~\bibnamefont
  {Laikhtman}},\ }\href {\doibase 10.1103/PhysRevB.67.235315} {\bibfield
  {journal} {\bibinfo  {journal} {Phys. Rev. B}\ }\textbf {\bibinfo {volume}
  {67}},\ \bibinfo {pages} {235315} (\bibinfo {year} {2003})}\BibitemShut
  {NoStop}%
\bibitem [{\citenamefont {Nikolaev}\ and\ \citenamefont
  {Portnoi}(2008)}]{Nikolaev2008}%
  \BibitemOpen
  \bibfield  {author} {\bibinfo {author} {\bibfnamefont {V.}~\bibnamefont
  {Nikolaev}}\ and\ \bibinfo {author} {\bibfnamefont {M.}~\bibnamefont
  {Portnoi}},\ }\href {\doibase 10.1016/j.spmi.2007.07.012} {\bibfield
  {journal} {\bibinfo  {journal} {Superlattices Microstruct.}\ }\textbf
  {\bibinfo {volume} {43}},\ \bibinfo {pages} {460} (\bibinfo {year}
  {2008})}\BibitemShut {NoStop}%
\bibitem [{\citenamefont {Lozovik}\ and\ \citenamefont
  {Berman}(1996)}]{Lozovik1996a}%
  \BibitemOpen
  \bibfield  {author} {\bibinfo {author} {\bibfnamefont {Y.~E.}\ \bibnamefont
  {Lozovik}}\ and\ \bibinfo {author} {\bibfnamefont {O.~L.}\ \bibnamefont
  {Berman}},\ }\href {\doibase 10.1134/1.567264} {\bibfield  {journal}
  {\bibinfo  {journal} {J. Exp. Theor. Phys. Lett.}\ }\textbf {\bibinfo
  {volume} {64}},\ \bibinfo {pages} {573} (\bibinfo {year} {1996})}\BibitemShut
  {NoStop}%
\bibitem [{\citenamefont {Koch}\ \emph {et~al.}(2003)\citenamefont {Koch},
  \citenamefont {Hoyer}, \citenamefont {Kira},\ and\ \citenamefont
  {Filinov}}]{Koch2003}%
  \BibitemOpen
  \bibfield  {author} {\bibinfo {author} {\bibfnamefont {S.~W.}\ \bibnamefont
  {Koch}}, \bibinfo {author} {\bibfnamefont {W.}~\bibnamefont {Hoyer}},
  \bibinfo {author} {\bibfnamefont {M.}~\bibnamefont {Kira}}, \ and\ \bibinfo
  {author} {\bibfnamefont {V.~S.}\ \bibnamefont {Filinov}},\ }\href {\doibase
  10.1002/pssb.200303153} {\bibfield  {journal} {\bibinfo  {journal} {Phys.
  Status Solidi}\ }\textbf {\bibinfo {volume} {238}},\ \bibinfo {pages} {404}
  (\bibinfo {year} {2003})}\BibitemShut {NoStop}%
\bibitem [{\citenamefont {Manzke}\ \emph {et~al.}(2012)\citenamefont {Manzke},
  \citenamefont {Semkat},\ and\ \citenamefont {Stolz}}]{Manzke2012}%
  \BibitemOpen
  \bibfield  {author} {\bibinfo {author} {\bibfnamefont {G.}~\bibnamefont
  {Manzke}}, \bibinfo {author} {\bibfnamefont {D.}~\bibnamefont {Semkat}}, \
  and\ \bibinfo {author} {\bibfnamefont {H.}~\bibnamefont {Stolz}},\ }\href
  {\doibase 10.1088/1367-2630/14/9/095002} {\bibfield  {journal} {\bibinfo
  {journal} {New J. Phys.}\ }\textbf {\bibinfo {volume} {14}},\ \bibinfo
  {pages} {095002} (\bibinfo {year} {2012})}\BibitemShut {NoStop}%
\bibitem [{\citenamefont {Chen}\ \emph {et~al.}(1987)\citenamefont {Chen},
  \citenamefont {Koteles}, \citenamefont {Elman},\ and\ \citenamefont
  {Armiento}}]{Chen1987}%
  \BibitemOpen
  \bibfield  {author} {\bibinfo {author} {\bibfnamefont {Y.~J.}\ \bibnamefont
  {Chen}}, \bibinfo {author} {\bibfnamefont {E.~S.}\ \bibnamefont {Koteles}},
  \bibinfo {author} {\bibfnamefont {B.~S.}\ \bibnamefont {Elman}}, \ and\
  \bibinfo {author} {\bibfnamefont {C.~A.}\ \bibnamefont {Armiento}},\ }\href
  {\doibase 10.1103/PhysRevB.36.4562} {\bibfield  {journal} {\bibinfo
  {journal} {Phys. Rev. B}\ }\textbf {\bibinfo {volume} {36}},\ \bibinfo
  {pages} {4562} (\bibinfo {year} {1987})}\BibitemShut {NoStop}%
\bibitem [{\citenamefont {Fisher}\ and\ \citenamefont
  {Hohenberg}(1988)}]{Fisher1988}%
  \BibitemOpen
  \bibfield  {author} {\bibinfo {author} {\bibfnamefont {D.}~\bibnamefont
  {Fisher}}\ and\ \bibinfo {author} {\bibfnamefont {P.}~\bibnamefont
  {Hohenberg}},\ }\href {\doibase 10.1103/PhysRevB.37.4936} {\bibfield
  {journal} {\bibinfo  {journal} {Phys. Rev. B}\ }\textbf {\bibinfo {volume}
  {37}},\ \bibinfo {pages} {4936} (\bibinfo {year} {1988})}\BibitemShut
  {NoStop}%
\bibitem [{\citenamefont {Hagn}\ \emph {et~al.}(1995)\citenamefont {Hagn},
  \citenamefont {Zrenner}, \citenamefont {B\"{o}hm},\ and\ \citenamefont
  {Weimann}}]{Hagn1995}%
  \BibitemOpen
  \bibfield  {author} {\bibinfo {author} {\bibfnamefont {M.}~\bibnamefont
  {Hagn}}, \bibinfo {author} {\bibfnamefont {A.}~\bibnamefont {Zrenner}},
  \bibinfo {author} {\bibfnamefont {G.}~\bibnamefont {B\"{o}hm}}, \ and\
  \bibinfo {author} {\bibfnamefont {G.}~\bibnamefont {Weimann}},\ }\href
  {\doibase 10.1063/1.114677} {\bibfield  {journal} {\bibinfo  {journal} {Appl.
  Phys. Lett.}\ }\textbf {\bibinfo {volume} {67}},\ \bibinfo {pages} {232}
  (\bibinfo {year} {1995})}\BibitemShut {NoStop}%
\bibitem [{\citenamefont {Kim}\ \emph {et~al.}(2005)\citenamefont {Kim},
  \citenamefont {Kim},\ and\ \citenamefont {Yoo}}]{Kim2005}%
  \BibitemOpen
  \bibfield  {author} {\bibinfo {author} {\bibfnamefont {J.~H.}\ \bibnamefont
  {Kim}}, \bibinfo {author} {\bibfnamefont {T.~W.}\ \bibnamefont {Kim}}, \ and\
  \bibinfo {author} {\bibfnamefont {K.~H.}\ \bibnamefont {Yoo}},\ }\href
  {\doibase 10.1016/j.apsusc.2004.07.032} {\bibfield  {journal} {\bibinfo
  {journal} {Appl. Surf. Sci.}\ }\textbf {\bibinfo {volume} {240}},\ \bibinfo
  {pages} {452} (\bibinfo {year} {2005})}\BibitemShut {NoStop}%
\bibitem [{\citenamefont {Remeika}\ \emph {et~al.}(2012)\citenamefont
  {Remeika}, \citenamefont {Fogler}, \citenamefont {Butov}, \citenamefont
  {Hanson},\ and\ \citenamefont {Gossard}}]{Remeika2012}%
  \BibitemOpen
  \bibfield  {author} {\bibinfo {author} {\bibfnamefont {M.}~\bibnamefont
  {Remeika}}, \bibinfo {author} {\bibfnamefont {M.~M.}\ \bibnamefont {Fogler}},
  \bibinfo {author} {\bibfnamefont {L.~V.}\ \bibnamefont {Butov}}, \bibinfo
  {author} {\bibfnamefont {M.}~\bibnamefont {Hanson}}, \ and\ \bibinfo {author}
  {\bibfnamefont {A.~C.}\ \bibnamefont {Gossard}},\ }\href {\doibase
  10.1063/1.3682302} {\bibfield  {journal} {\bibinfo  {journal} {Appl. Phys.
  Lett.}\ }\textbf {\bibinfo {volume} {100}},\ \bibinfo {pages} {061103}
  (\bibinfo {year} {2012})}\BibitemShut {NoStop}%
\bibitem [{\citenamefont {Winbow}\ \emph {et~al.}(2011)\citenamefont {Winbow},
  \citenamefont {Leonard}, \citenamefont {Remeika}, \citenamefont {Kuznetsova},
  \citenamefont {High}, \citenamefont {Hammack}, \citenamefont {Butov},
  \citenamefont {Wilkes}, \citenamefont {Guenther}, \citenamefont {Ivanov},
  \citenamefont {Hanson},\ and\ \citenamefont {Gossard}}]{Winbow2011}%
  \BibitemOpen
  \bibfield  {author} {\bibinfo {author} {\bibfnamefont {A.~G.}\ \bibnamefont
  {Winbow}}, \bibinfo {author} {\bibfnamefont {J.~R.}\ \bibnamefont {Leonard}},
  \bibinfo {author} {\bibfnamefont {M.}~\bibnamefont {Remeika}}, \bibinfo
  {author} {\bibfnamefont {Y.~Y.}\ \bibnamefont {Kuznetsova}}, \bibinfo
  {author} {\bibfnamefont {A.~A.}\ \bibnamefont {High}}, \bibinfo {author}
  {\bibfnamefont {A.~T.}\ \bibnamefont {Hammack}}, \bibinfo {author}
  {\bibfnamefont {L.~V.}\ \bibnamefont {Butov}}, \bibinfo {author}
  {\bibfnamefont {J.}~\bibnamefont {Wilkes}}, \bibinfo {author} {\bibfnamefont
  {A.~A.}\ \bibnamefont {Guenther}}, \bibinfo {author} {\bibfnamefont {A.~L.}\
  \bibnamefont {Ivanov}}, \bibinfo {author} {\bibfnamefont {M.}~\bibnamefont
  {Hanson}}, \ and\ \bibinfo {author} {\bibfnamefont {A.~C.}\ \bibnamefont
  {Gossard}},\ }\href {\doibase 10.1103/PhysRevLett.106.196806} {\bibfield
  {journal} {\bibinfo  {journal} {Phys. Rev. Lett.}\ }\textbf {\bibinfo
  {volume} {106}},\ \bibinfo {pages} {196806} (\bibinfo {year}
  {2011})}\BibitemShut {NoStop}%
\bibitem [{\citenamefont {Butov}\ and\ \citenamefont
  {Filin}(1998)}]{Butov1998}%
  \BibitemOpen
  \bibfield  {author} {\bibinfo {author} {\bibfnamefont {L.~V.}\ \bibnamefont
  {Butov}}\ and\ \bibinfo {author} {\bibfnamefont {A.}~\bibnamefont {Filin}},\
  }\href {\doibase 10.1103/PhysRevB.58.1980} {\bibfield  {journal} {\bibinfo
  {journal} {Phys. Rev. B}\ }\textbf {\bibinfo {volume} {58}},\ \bibinfo
  {pages} {1980} (\bibinfo {year} {1998})}\BibitemShut {NoStop}%
\bibitem [{\citenamefont {Butov}\ \emph {et~al.}(1994)\citenamefont {Butov},
  \citenamefont {Zrenner}, \citenamefont {Abstreiter}, \citenamefont
  {B\"{o}hm},\ and\ \citenamefont {Weimann}}]{Butov1994}%
  \BibitemOpen
  \bibfield  {author} {\bibinfo {author} {\bibfnamefont {L.~V.}\ \bibnamefont
  {Butov}}, \bibinfo {author} {\bibfnamefont {A.}~\bibnamefont {Zrenner}},
  \bibinfo {author} {\bibfnamefont {G.}~\bibnamefont {Abstreiter}}, \bibinfo
  {author} {\bibfnamefont {G.}~\bibnamefont {B\"{o}hm}}, \ and\ \bibinfo
  {author} {\bibfnamefont {G.}~\bibnamefont {Weimann}},\ }\href {\doibase
  10.1103/PhysRevLett.73.304} {\bibfield  {journal} {\bibinfo  {journal} {Phys.
  Rev. Lett.}\ }\textbf {\bibinfo {volume} {73}},\ \bibinfo {pages} {304}
  (\bibinfo {year} {1994})}\BibitemShut {NoStop}%
\bibitem [{\citenamefont {Larionov}\ and\ \citenamefont
  {Timofeev}(2001)}]{Larionov2001}%
  \BibitemOpen
  \bibfield  {author} {\bibinfo {author} {\bibfnamefont {A.~V.}\ \bibnamefont
  {Larionov}}\ and\ \bibinfo {author} {\bibfnamefont {V.~B.}\ \bibnamefont
  {Timofeev}},\ }\href {\doibase 10.1134/1.1374266} {\bibfield  {journal}
  {\bibinfo  {journal} {J. Exp. Theor. Phys. Lett.}\ }\textbf {\bibinfo
  {volume} {73}},\ \bibinfo {pages} {301} (\bibinfo {year} {2001})}\BibitemShut
  {NoStop}%
\bibitem [{\citenamefont {Butov}\ \emph {et~al.}(2002)\citenamefont {Butov},
  \citenamefont {Lai}, \citenamefont {Ivanov}, \citenamefont {Gossard},\ and\
  \citenamefont {Chemla}}]{Butov2002a}%
  \BibitemOpen
  \bibfield  {author} {\bibinfo {author} {\bibfnamefont {L.~V.}\ \bibnamefont
  {Butov}}, \bibinfo {author} {\bibfnamefont {C.~W.}\ \bibnamefont {Lai}},
  \bibinfo {author} {\bibfnamefont {A.~L.}\ \bibnamefont {Ivanov}}, \bibinfo
  {author} {\bibfnamefont {A.~C.}\ \bibnamefont {Gossard}}, \ and\ \bibinfo
  {author} {\bibfnamefont {D.~S.}\ \bibnamefont {Chemla}},\ }\href {\doibase
  10.1038/417047a} {\bibfield  {journal} {\bibinfo  {journal} {Nature}\
  }\textbf {\bibinfo {volume} {417}},\ \bibinfo {pages} {47} (\bibinfo {year}
  {2002})}\BibitemShut {NoStop}%
\bibitem [{\citenamefont {Christmann}\ \emph {et~al.}(2011)\citenamefont
  {Christmann}, \citenamefont {Askitopoulos}, \citenamefont {Deligeorgis},
  \citenamefont {Hatzopoulos}, \citenamefont {Tsintzos}, \citenamefont
  {Savvidis},\ and\ \citenamefont {Baumberg}}]{Christmann2011}%
  \BibitemOpen
  \bibfield  {author} {\bibinfo {author} {\bibfnamefont {G.}~\bibnamefont
  {Christmann}}, \bibinfo {author} {\bibfnamefont {A.}~\bibnamefont
  {Askitopoulos}}, \bibinfo {author} {\bibfnamefont {G.}~\bibnamefont
  {Deligeorgis}}, \bibinfo {author} {\bibfnamefont {Z.}~\bibnamefont
  {Hatzopoulos}}, \bibinfo {author} {\bibfnamefont {S.~I.}\ \bibnamefont
  {Tsintzos}}, \bibinfo {author} {\bibfnamefont {P.~G.}\ \bibnamefont
  {Savvidis}}, \ and\ \bibinfo {author} {\bibfnamefont {J.~J.}\ \bibnamefont
  {Baumberg}},\ }\href {\doibase 10.1063/1.3559909} {\bibfield  {journal}
  {\bibinfo  {journal} {Appl. Phys. Lett.}\ }\textbf {\bibinfo {volume} {98}},\
  \bibinfo {pages} {081111} (\bibinfo {year} {2011})}\BibitemShut {NoStop}%
\bibitem [{\citenamefont {Luo}\ \emph {et~al.}(1994)\citenamefont {Luo},
  \citenamefont {Planken}, \citenamefont {Brener}, \citenamefont {Roskos},\
  and\ \citenamefont {Nuss}}]{Luo1994}%
  \BibitemOpen
  \bibfield  {author} {\bibinfo {author} {\bibfnamefont {M.~S.~C.}\
  \bibnamefont {Luo}}, \bibinfo {author} {\bibfnamefont {P.~C.~M.}\
  \bibnamefont {Planken}}, \bibinfo {author} {\bibfnamefont {I.}~\bibnamefont
  {Brener}}, \bibinfo {author} {\bibfnamefont {H.~G.}\ \bibnamefont {Roskos}},
  \ and\ \bibinfo {author} {\bibfnamefont {M.~C.}\ \bibnamefont {Nuss}},\
  }\href {\doibase 10.1109/3.299473} {\bibfield  {journal} {\bibinfo  {journal}
  {IEEE J. Quantum Electron.}\ }\textbf {\bibinfo {volume} {30}},\ \bibinfo
  {pages} {1478} (\bibinfo {year} {1994})}\BibitemShut {NoStop}%
\bibitem [{\citenamefont {Daveau}\ \emph {et~al.}(2015)\citenamefont {Daveau},
  \citenamefont {Tighineanu}, \citenamefont {Lodahl},\ and\ \citenamefont
  {Stobbe}}]{Daveaud2015}%
  \BibitemOpen
  \bibfield  {author} {\bibinfo {author} {\bibfnamefont {R.~S.}\ \bibnamefont
  {Daveau}}, \bibinfo {author} {\bibfnamefont {P.}~\bibnamefont {Tighineanu}},
  \bibinfo {author} {\bibfnamefont {P.}~\bibnamefont {Lodahl}}, \ and\ \bibinfo
  {author} {\bibfnamefont {S.}~\bibnamefont {Stobbe}},\ }\href {\doibase
  10.1364/OE.23.025340} {\bibfield  {journal} {\bibinfo  {journal} {Opt.
  Express}\ }\textbf {\bibinfo {volume} {23}},\ \bibinfo {pages} {25340}
  (\bibinfo {year} {2015})}\BibitemShut {NoStop}%
\bibitem [{\citenamefont {Burau}\ \emph {et~al.}(2010)\citenamefont {Burau},
  \citenamefont {Manzke}, \citenamefont {Kieseling}, \citenamefont {Stolz},
  \citenamefont {Reuter},\ and\ \citenamefont {Wieck}}]{Burau2010}%
  \BibitemOpen
  \bibfield  {author} {\bibinfo {author} {\bibfnamefont {G.~K.~G.}\
  \bibnamefont {Burau}}, \bibinfo {author} {\bibfnamefont {G.}~\bibnamefont
  {Manzke}}, \bibinfo {author} {\bibfnamefont {F.}~\bibnamefont {Kieseling}},
  \bibinfo {author} {\bibfnamefont {H.}~\bibnamefont {Stolz}}, \bibinfo
  {author} {\bibfnamefont {D.}~\bibnamefont {Reuter}}, \ and\ \bibinfo {author}
  {\bibfnamefont {A.}~\bibnamefont {Wieck}},\ }\href {\doibase
  10.1088/1742-6596/210/1/012017} {\bibfield  {journal} {\bibinfo  {journal}
  {J. Phys. Conf. Ser.}\ }\textbf {\bibinfo {volume} {210}},\ \bibinfo {pages}
  {012017} (\bibinfo {year} {2010})}\BibitemShut {NoStop}%
\bibitem [{\citenamefont {Rice}(1978)}]{Rice1978}%
  \BibitemOpen
  \bibfield  {author} {\bibinfo {author} {\bibfnamefont {T.}~\bibnamefont
  {Rice}},\ }\href {\doibase 10.1016/S0081-1947(08)60438-5} {\bibfield
  {journal} {\bibinfo  {journal} {Solid State Phys.}\ }\textbf {\bibinfo
  {volume} {32}},\ \bibinfo {pages} {1} (\bibinfo {year} {1978})}\BibitemShut
  {NoStop}%
\bibitem [{\citenamefont {Szymanska}\ and\ \citenamefont
  {Littlewood}(2003)}]{Szymanska2003}%
  \BibitemOpen
  \bibfield  {author} {\bibinfo {author} {\bibfnamefont {M.~H.}\ \bibnamefont
  {Szymanska}}\ and\ \bibinfo {author} {\bibfnamefont {P.~B.}\ \bibnamefont
  {Littlewood}},\ }\href {\doibase 10.1103/PhysRevB.67.193305} {\bibfield
  {journal} {\bibinfo  {journal} {Phys. Rev. B}\ }\textbf {\bibinfo {volume}
  {67}},\ \bibinfo {pages} {193305} (\bibinfo {year} {2003})}\BibitemShut
  {NoStop}%
\bibitem [{\citenamefont {Sivalertporn}\ \emph {et~al.}(2012)\citenamefont
  {Sivalertporn}, \citenamefont {Mouchliadis}, \citenamefont {Ivanov},
  \citenamefont {Philp},\ and\ \citenamefont {Muljarov}}]{Sivalertporn2012}%
  \BibitemOpen
  \bibfield  {author} {\bibinfo {author} {\bibfnamefont {K.}~\bibnamefont
  {Sivalertporn}}, \bibinfo {author} {\bibfnamefont {L.}~\bibnamefont
  {Mouchliadis}}, \bibinfo {author} {\bibfnamefont {A.~L.}\ \bibnamefont
  {Ivanov}}, \bibinfo {author} {\bibfnamefont {R.}~\bibnamefont {Philp}}, \
  and\ \bibinfo {author} {\bibfnamefont {E.~A.}\ \bibnamefont {Muljarov}},\
  }\href {\doibase 10.1103/PhysRevB.85.045207} {\bibfield  {journal} {\bibinfo
  {journal} {Phys. Rev. B}\ }\textbf {\bibinfo {volume} {85}},\ \bibinfo
  {pages} {045207} (\bibinfo {year} {2012})}\BibitemShut {NoStop}%
\bibitem [{\citenamefont {Huml\'{\i}\v{c}ek}\ \emph {et~al.}(1993)\citenamefont
  {Huml\'{\i}\v{c}ek}, \citenamefont {Schmidt}, \citenamefont
  {Bo\v{c}\'{a}nek}, \citenamefont {\v{S}vehla},\ and\ \citenamefont
  {Ploog}}]{Humlicek1993}%
  \BibitemOpen
  \bibfield  {author} {\bibinfo {author} {\bibfnamefont {J.}~\bibnamefont
  {Huml\'{\i}\v{c}ek}}, \bibinfo {author} {\bibfnamefont {E.}~\bibnamefont
  {Schmidt}}, \bibinfo {author} {\bibfnamefont {L.}~\bibnamefont
  {Bo\v{c}\'{a}nek}}, \bibinfo {author} {\bibfnamefont {R.}~\bibnamefont
  {\v{S}vehla}}, \ and\ \bibinfo {author} {\bibfnamefont {K.}~\bibnamefont
  {Ploog}},\ }\href {\doibase 10.1103/PhysRevB.48.5241} {\bibfield  {journal}
  {\bibinfo  {journal} {Phys. Rev. B}\ }\textbf {\bibinfo {volume} {48}},\
  \bibinfo {pages} {5241} (\bibinfo {year} {1993})}\BibitemShut {NoStop}%
\bibitem [{\citenamefont {Schindler}\ and\ \citenamefont
  {Zimmermann}(2008)}]{Schindler2008}%
  \BibitemOpen
  \bibfield  {author} {\bibinfo {author} {\bibfnamefont {C.}~\bibnamefont
  {Schindler}}\ and\ \bibinfo {author} {\bibfnamefont {R.}~\bibnamefont
  {Zimmermann}},\ }\href {\doibase 10.1103/PhysRevB.78.045313} {\bibfield
  {journal} {\bibinfo  {journal} {Phys. Rev. B}\ }\textbf {\bibinfo {volume}
  {78}},\ \bibinfo {pages} {045313} (\bibinfo {year} {2008})}\BibitemShut
  {NoStop}%
\bibitem [{\citenamefont {Ben-Tabou~de Leon}\ and\ \citenamefont
  {Laikhtman}(2001)}]{Ben-Taboude-Leon2001}%
  \BibitemOpen
  \bibfield  {author} {\bibinfo {author} {\bibfnamefont {S.}~\bibnamefont
  {Ben-Tabou~de Leon}}\ and\ \bibinfo {author} {\bibfnamefont {B.}~\bibnamefont
  {Laikhtman}},\ }\href {\doibase 10.1103/PhysRevB.63.125306} {\bibfield
  {journal} {\bibinfo  {journal} {Phys. Rev. B}\ }\textbf {\bibinfo {volume}
  {63}},\ \bibinfo {pages} {125306} (\bibinfo {year} {2001})}\BibitemShut
  {NoStop}%
\bibitem [{\citenamefont {\v{C}erne}\ \emph {et~al.}(1996)\citenamefont
  {\v{C}erne}, \citenamefont {Kono}, \citenamefont {Sherwin}, \citenamefont
  {Sundaram}, \citenamefont {Gossard},\ and\ \citenamefont
  {Bauer}}]{Cerne1996}%
  \BibitemOpen
  \bibfield  {author} {\bibinfo {author} {\bibfnamefont {J.}~\bibnamefont
  {\v{C}erne}}, \bibinfo {author} {\bibfnamefont {J.}~\bibnamefont {Kono}},
  \bibinfo {author} {\bibfnamefont {M.~S.}\ \bibnamefont {Sherwin}}, \bibinfo
  {author} {\bibfnamefont {M.}~\bibnamefont {Sundaram}}, \bibinfo {author}
  {\bibfnamefont {A.~C.}\ \bibnamefont {Gossard}}, \ and\ \bibinfo {author}
  {\bibfnamefont {G.~E.~W.}\ \bibnamefont {Bauer}},\ }\href {\doibase
  10.1103/PhysRevLett.77.1131} {\bibfield  {journal} {\bibinfo  {journal}
  {Phys. Rev. Lett.}\ }\textbf {\bibinfo {volume} {77}},\ \bibinfo {pages}
  {1131} (\bibinfo {year} {1996})}\BibitemShut {NoStop}%
\bibitem [{\citenamefont {Snoke}(2008)}]{Snoke2008}%
  \BibitemOpen
  \bibfield  {author} {\bibinfo {author} {\bibfnamefont {D.}~\bibnamefont
  {Snoke}},\ }\href {\doibase 10.1016/j.ssc.2008.01.012} {\bibfield  {journal}
  {\bibinfo  {journal} {Solid State Commun.}\ }\textbf {\bibinfo {volume}
  {146}},\ \bibinfo {pages} {73} (\bibinfo {year} {2008})}\BibitemShut
  {NoStop}%
\bibitem [{emp()}]{empty}%
  \BibitemOpen
  \href@noop {} {\ }\BibitemShut {NoStop}%
\bibitem [{\citenamefont {Haug}\ and\ \citenamefont
  {Schmitt-Rink}(1984)}]{Haug1984}%
  \BibitemOpen
  \bibfield  {author} {\bibinfo {author} {\bibfnamefont {H.}~\bibnamefont
  {Haug}}\ and\ \bibinfo {author} {\bibfnamefont {S.}~\bibnamefont
  {Schmitt-Rink}},\ }\href {\doibase 10.1016/0079-6727(84)90026-0} {\bibfield
  {journal} {\bibinfo  {journal} {Prog. Quantum Electron.}\ }\textbf {\bibinfo
  {volume} {9}},\ \bibinfo {pages} {3} (\bibinfo {year} {1984})}\BibitemShut
  {NoStop}%
\bibitem [{\citenamefont {Zimmermann}(1988)}]{Zimmermann1988}%
  \BibitemOpen
  \bibfield  {author} {\bibinfo {author} {\bibfnamefont {R.}~\bibnamefont
  {Zimmermann}},\ }\href {\doibase 10.1002/pssb.2221460140} {\bibfield
  {journal} {\bibinfo  {journal} {Phys. Status Solidi}\ }\textbf {\bibinfo
  {volume} {146}},\ \bibinfo {pages} {371} (\bibinfo {year}
  {1988})}\BibitemShut {NoStop}%
\bibitem [{\citenamefont {Miller}\ \emph {et~al.}(1985)\citenamefont {Miller},
  \citenamefont {Chemla}, \citenamefont {Damen}, \citenamefont {Gossard},
  \citenamefont {Wiegmann}, \citenamefont {Wood},\ and\ \citenamefont
  {Burrus}}]{Miller1985}%
  \BibitemOpen
  \bibfield  {author} {\bibinfo {author} {\bibfnamefont {D.~A.~B.}\
  \bibnamefont {Miller}}, \bibinfo {author} {\bibfnamefont {D.~S.}\
  \bibnamefont {Chemla}}, \bibinfo {author} {\bibfnamefont {T.~C.}\
  \bibnamefont {Damen}}, \bibinfo {author} {\bibfnamefont {A.~C.}\ \bibnamefont
  {Gossard}}, \bibinfo {author} {\bibfnamefont {W.}~\bibnamefont {Wiegmann}},
  \bibinfo {author} {\bibfnamefont {T.~H.}\ \bibnamefont {Wood}}, \ and\
  \bibinfo {author} {\bibfnamefont {C.~A.}\ \bibnamefont {Burrus}},\ }\href
  {\doibase 10.1103/PhysRevB.32.1043} {\bibfield  {journal} {\bibinfo
  {journal} {Phys. Rev. B}\ }\textbf {\bibinfo {volume} {32}},\ \bibinfo
  {pages} {1043} (\bibinfo {year} {1985})}\BibitemShut {NoStop}%
\bibitem [{\citenamefont {Vurgaftman}\ \emph {et~al.}(2001)\citenamefont
  {Vurgaftman}, \citenamefont {Meyer},\ and\ \citenamefont
  {Ram-Mohan}}]{Vurgaftman2001}%
  \BibitemOpen
  \bibfield  {author} {\bibinfo {author} {\bibfnamefont {I.}~\bibnamefont
  {Vurgaftman}}, \bibinfo {author} {\bibfnamefont {J.~R.}\ \bibnamefont
  {Meyer}}, \ and\ \bibinfo {author} {\bibfnamefont {L.~R.}\ \bibnamefont
  {Ram-Mohan}},\ }\href {\doibase 10.1063/1.1368156} {\bibfield  {journal}
  {\bibinfo  {journal} {J. Appl. Phys.}\ }\textbf {\bibinfo {volume} {89}},\
  \bibinfo {pages} {5815} (\bibinfo {year} {2001})}\BibitemShut {NoStop}%
\bibitem [{\citenamefont {Andrews}\ \emph {et~al.}(1988)\citenamefont
  {Andrews}, \citenamefont {Murray}, \citenamefont {Davies},\ and\
  \citenamefont {Kerr}}]{Andrews1988}%
  \BibitemOpen
  \bibfield  {author} {\bibinfo {author} {\bibfnamefont {S.~R.}\ \bibnamefont
  {Andrews}}, \bibinfo {author} {\bibfnamefont {C.~M.}\ \bibnamefont {Murray}},
  \bibinfo {author} {\bibfnamefont {R.~A.}\ \bibnamefont {Davies}}, \ and\
  \bibinfo {author} {\bibfnamefont {T.~M.}\ \bibnamefont {Kerr}},\ }\href
  {\doibase 10.1103/PhysRevB.37.8198} {\bibfield  {journal} {\bibinfo
  {journal} {Phys. Rev. B}\ }\textbf {\bibinfo {volume} {37}},\ \bibinfo
  {pages} {8198} (\bibinfo {year} {1988})}\BibitemShut {NoStop}%
\bibitem [{\citenamefont {Butov}\ \emph {et~al.}(2001)\citenamefont {Butov},
  \citenamefont {Ivanov}, \citenamefont {Imamoglu}, \citenamefont {Littlewood},
  \citenamefont {Shashkin}, \citenamefont {Dolgopolov}, \citenamefont
  {Campman},\ and\ \citenamefont {Gossard}}]{Butov2001}%
  \BibitemOpen
  \bibfield  {author} {\bibinfo {author} {\bibfnamefont {L.~V.}\ \bibnamefont
  {Butov}}, \bibinfo {author} {\bibfnamefont {A.~L.}\ \bibnamefont {Ivanov}},
  \bibinfo {author} {\bibfnamefont {A.}~\bibnamefont {Imamoglu}}, \bibinfo
  {author} {\bibfnamefont {P.~B.}\ \bibnamefont {Littlewood}}, \bibinfo
  {author} {\bibfnamefont {A.~A.}\ \bibnamefont {Shashkin}}, \bibinfo {author}
  {\bibfnamefont {V.~T.}\ \bibnamefont {Dolgopolov}}, \bibinfo {author}
  {\bibfnamefont {K.~L.}\ \bibnamefont {Campman}}, \ and\ \bibinfo {author}
  {\bibfnamefont {A.~C.}\ \bibnamefont {Gossard}},\ }\href {\doibase
  10.1103/PhysRevLett.86.5608} {\bibfield  {journal} {\bibinfo  {journal}
  {Phys. Rev. Lett.}\ }\textbf {\bibinfo {volume} {86}},\ \bibinfo {pages}
  {5608} (\bibinfo {year} {2001})}\BibitemShut {NoStop}%
\bibitem [{\citenamefont {Amo}\ \emph {et~al.}(2007)\citenamefont {Amo},
  \citenamefont {Marti`n}, \citenamefont {Vi\~{n}a}, \citenamefont {Toropov},\
  and\ \citenamefont {Zhuravlev}}]{Amo2007}%
  \BibitemOpen
  \bibfield  {author} {\bibinfo {author} {\bibfnamefont {A.}~\bibnamefont
  {Amo}}, \bibinfo {author} {\bibfnamefont {M.~D.}\ \bibnamefont {Marti`n}},
  \bibinfo {author} {\bibfnamefont {L.}~\bibnamefont {Vi\~{n}a}}, \bibinfo
  {author} {\bibfnamefont {A.~I.}\ \bibnamefont {Toropov}}, \ and\ \bibinfo
  {author} {\bibfnamefont {K.~S.}\ \bibnamefont {Zhuravlev}},\ }\href {\doibase
  10.1063/1.2722786} {\bibfield  {journal} {\bibinfo  {journal} {J. Appl.
  Phys.}\ }\textbf {\bibinfo {volume} {101}},\ \bibinfo {pages} {081717}
  (\bibinfo {year} {2007})}\BibitemShut {NoStop}%
\bibitem [{\citenamefont {Shahmohammadi}\ \emph {et~al.}(2014)\citenamefont
  {Shahmohammadi}, \citenamefont {Jacopin}, \citenamefont {Rossbach},
  \citenamefont {Levrat}, \citenamefont {Feltin}, \citenamefont {Carlin},
  \citenamefont {Gani\`{e}re}, \citenamefont {Butt\'{e}}, \citenamefont
  {Grandjean},\ and\ \citenamefont {Deveaud}}]{Shahmohammadi2014}%
  \BibitemOpen
  \bibfield  {author} {\bibinfo {author} {\bibfnamefont {M.}~\bibnamefont
  {Shahmohammadi}}, \bibinfo {author} {\bibfnamefont {G.}~\bibnamefont
  {Jacopin}}, \bibinfo {author} {\bibfnamefont {G.}~\bibnamefont {Rossbach}},
  \bibinfo {author} {\bibfnamefont {J.}~\bibnamefont {Levrat}}, \bibinfo
  {author} {\bibfnamefont {E.}~\bibnamefont {Feltin}}, \bibinfo {author}
  {\bibfnamefont {J.-F.}\ \bibnamefont {Carlin}}, \bibinfo {author}
  {\bibfnamefont {J.-D.}\ \bibnamefont {Gani\`{e}re}}, \bibinfo {author}
  {\bibfnamefont {R.}~\bibnamefont {Butt\'{e}}}, \bibinfo {author}
  {\bibfnamefont {N.}~\bibnamefont {Grandjean}}, \ and\ \bibinfo {author}
  {\bibfnamefont {B.}~\bibnamefont {Deveaud}},\ }\href {\doibase
  10.1038/ncomms6251} {\bibfield  {journal} {\bibinfo  {journal} {Nat.
  Commun.}\ }\textbf {\bibinfo {volume} {5}},\ \bibinfo {pages} {5251}
  (\bibinfo {year} {2014})}\BibitemShut {NoStop}%
\bibitem [{\citenamefont {High}\ \emph {et~al.}(2009)\citenamefont {High},
  \citenamefont {Hammack}, \citenamefont {Butov}, \citenamefont {Mouchliadis},
  \citenamefont {Ivanov}, \citenamefont {Hanson},\ and\ \citenamefont
  {Gossard}}]{High2009}%
  \BibitemOpen
  \bibfield  {author} {\bibinfo {author} {\bibfnamefont {A.~A.}\ \bibnamefont
  {High}}, \bibinfo {author} {\bibfnamefont {A.~T.}\ \bibnamefont {Hammack}},
  \bibinfo {author} {\bibfnamefont {L.~V.}\ \bibnamefont {Butov}}, \bibinfo
  {author} {\bibfnamefont {L.}~\bibnamefont {Mouchliadis}}, \bibinfo {author}
  {\bibfnamefont {A.~L.}\ \bibnamefont {Ivanov}}, \bibinfo {author}
  {\bibfnamefont {M.}~\bibnamefont {Hanson}}, \ and\ \bibinfo {author}
  {\bibfnamefont {A.~C.}\ \bibnamefont {Gossard}},\ }\href {\doibase
  10.1021/nl900605b} {\bibfield  {journal} {\bibinfo  {journal} {Nano Lett.}\
  }\textbf {\bibinfo {volume} {9}},\ \bibinfo {pages} {2094} (\bibinfo {year}
  {2009})}\BibitemShut {NoStop}%
\bibitem [{\citenamefont {Sugakov}(2014)}]{Sugakov2014}%
  \BibitemOpen
  \bibfield  {author} {\bibinfo {author} {\bibfnamefont {V.~I.}\ \bibnamefont
  {Sugakov}},\ }\href {\doibase 10.5488/CMP.17.33702} {\bibfield  {journal}
  {\bibinfo  {journal} {Condens. Matter Phys.}\ }\textbf {\bibinfo {volume}
  {17}},\ \bibinfo {pages} {33702} (\bibinfo {year} {2014})}\BibitemShut
  {NoStop}%
\bibitem [{\citenamefont {Hoger}\ \emph {et~al.}(1984)\citenamefont {Hoger},
  \citenamefont {Gobel}, \citenamefont {Kuhl}, \citenamefont {Ploog},\ and\
  \citenamefont {Quiesser}}]{Hoger1984}%
  \BibitemOpen
  \bibfield  {author} {\bibinfo {author} {\bibfnamefont {R.}~\bibnamefont
  {Hoger}}, \bibinfo {author} {\bibfnamefont {E.~O.}\ \bibnamefont {Gobel}},
  \bibinfo {author} {\bibfnamefont {J.}~\bibnamefont {Kuhl}}, \bibinfo {author}
  {\bibfnamefont {K.}~\bibnamefont {Ploog}}, \ and\ \bibinfo {author}
  {\bibfnamefont {H.~J.}\ \bibnamefont {Quiesser}},\ }\href {\doibase
  10.1088/0022-3719/17/34/002} {\bibfield  {journal} {\bibinfo  {journal} {J.
  Phys. C Solid State Phys.}\ }\textbf {\bibinfo {volume} {17}},\ \bibinfo
  {pages} {L905} (\bibinfo {year} {1984})}\BibitemShut {NoStop}%
\bibitem [{\citenamefont {Leosson}\ \emph {et~al.}(2000)\citenamefont
  {Leosson}, \citenamefont {Jensen}, \citenamefont {Langbein},\ and\
  \citenamefont {Hvam}}]{Leosson2000}%
  \BibitemOpen
  \bibfield  {author} {\bibinfo {author} {\bibfnamefont {K.}~\bibnamefont
  {Leosson}}, \bibinfo {author} {\bibfnamefont {J.~R.}\ \bibnamefont {Jensen}},
  \bibinfo {author} {\bibfnamefont {W.}~\bibnamefont {Langbein}}, \ and\
  \bibinfo {author} {\bibfnamefont {J.~M.}\ \bibnamefont {Hvam}},\ }\href
  {\doibase 10.1103/PhysRevB.61.10322} {\bibfield  {journal} {\bibinfo
  {journal} {Phys. Rev. B}\ }\textbf {\bibinfo {volume} {61}},\ \bibinfo
  {pages} {10322} (\bibinfo {year} {2000})}\BibitemShut {NoStop}%
\bibitem [{\citenamefont {Larionov}\ \emph {et~al.}(2002)\citenamefont
  {Larionov}, \citenamefont {Timofeev}, \citenamefont {Ni}, \citenamefont
  {Dubonos}, \citenamefont {Hvam},\ and\ \citenamefont
  {Soerensen}}]{Larionov2002}%
  \BibitemOpen
  \bibfield  {author} {\bibinfo {author} {\bibfnamefont {A.~V.}\ \bibnamefont
  {Larionov}}, \bibinfo {author} {\bibfnamefont {V.~B.}\ \bibnamefont
  {Timofeev}}, \bibinfo {author} {\bibfnamefont {P.~A.}\ \bibnamefont {Ni}},
  \bibinfo {author} {\bibfnamefont {S.~V.}\ \bibnamefont {Dubonos}}, \bibinfo
  {author} {\bibfnamefont {I.}~\bibnamefont {Hvam}}, \ and\ \bibinfo {author}
  {\bibfnamefont {K.}~\bibnamefont {Soerensen}},\ }\href {\doibase
  10.1134/1.1500724} {\bibfield  {journal} {\bibinfo  {journal} {J. Exp. Theor.
  Phys. Lett.}\ }\textbf {\bibinfo {volume} {75}},\ \bibinfo {pages} {570}
  (\bibinfo {year} {2002})}\BibitemShut {NoStop}%
\bibitem [{\citenamefont {Dremin}\ \emph {et~al.}(2002)\citenamefont {Dremin},
  \citenamefont {Timofeev}, \citenamefont {Larionov}, \citenamefont {Hvam},\
  and\ \citenamefont {Soerensen}}]{Dremin2002}%
  \BibitemOpen
  \bibfield  {author} {\bibinfo {author} {\bibfnamefont {A.~A.}\ \bibnamefont
  {Dremin}}, \bibinfo {author} {\bibfnamefont {V.~B.}\ \bibnamefont
  {Timofeev}}, \bibinfo {author} {\bibfnamefont {A.~V.}\ \bibnamefont
  {Larionov}}, \bibinfo {author} {\bibfnamefont {J.}~\bibnamefont {Hvam}}, \
  and\ \bibinfo {author} {\bibfnamefont {K.}~\bibnamefont {Soerensen}},\ }\href
  {\doibase 10.1134/1.1528700} {\bibfield  {journal} {\bibinfo  {journal} {J.
  Exp. Theor. Phys. Lett.}\ }\textbf {\bibinfo {volume} {76}},\ \bibinfo
  {pages} {450} (\bibinfo {year} {2002})}\BibitemShut {NoStop}%
\bibitem [{\citenamefont {Cui}\ \emph {et~al.}(1996)\citenamefont {Cui},
  \citenamefont {Ding}, \citenamefont {Lee}, \citenamefont {Veliadis},
  \citenamefont {Khurgin}, \citenamefont {Li}, \citenamefont {Reynolds},\ and\
  \citenamefont {Grata}}]{Cui1996}%
  \BibitemOpen
  \bibfield  {author} {\bibinfo {author} {\bibfnamefont {A.~G.}\ \bibnamefont
  {Cui}}, \bibinfo {author} {\bibfnamefont {Y.~J.}\ \bibnamefont {Ding}},
  \bibinfo {author} {\bibfnamefont {S.~J.}\ \bibnamefont {Lee}}, \bibinfo
  {author} {\bibfnamefont {J.~V.~D.}\ \bibnamefont {Veliadis}}, \bibinfo
  {author} {\bibfnamefont {J.~B.}\ \bibnamefont {Khurgin}}, \bibinfo {author}
  {\bibfnamefont {S.}~\bibnamefont {Li}}, \bibinfo {author} {\bibfnamefont
  {D.~C.}\ \bibnamefont {Reynolds}}, \ and\ \bibinfo {author} {\bibfnamefont
  {J.}~\bibnamefont {Grata}},\ }\href {\doibase 10.1364/JOSAB.13.000536}
  {\bibfield  {journal} {\bibinfo  {journal} {J. Opt. Soc. Am. B}\ }\textbf
  {\bibinfo {volume} {13}},\ \bibinfo {pages} {536} (\bibinfo {year}
  {1996})}\BibitemShut {NoStop}%
\end{thebibliography}
%

\end{document}